\documentclass[twocolumn]{emulateapj}
\usepackage{graphicx}
\usepackage{float}
\usepackage{mathrsfs}	
\usepackage{amsmath}
\usepackage{color}
\usepackage{multirow}
\usepackage{subfigure}
\usepackage{longtable}
\usepackage{blindtext}
\usepackage{lineno}
\newcommand{\rui}[1]{{\color{black}{#1}}}

\newcommand{\nic}[1]{{\color{black}{#1}}}
\def\lsim{\mathrel{\rlap{\lower3.5pt\hbox{\hskip0.5pt$\sim$}}
    \raise0.5pt\hbox{$<$}}} 
\begin{document}
\title{
GAlaxy Light profile convolutional neural NETworks (GaLNets). I. fast and accurate structural parameters for billion galaxy samples 
}

\author{\mbox{R. Li\altaffilmark{1,2,3}}}
\author{\mbox{N. R. Napolitano\altaffilmark{2,4,5}}}
\author{\mbox{N. Roy\altaffilmark{6}}}
\author{\mbox{C. Tortora\altaffilmark{5}}}
\author{\mbox{F. La Barbera\altaffilmark{5}}}
\author{\mbox{A. sonnenfeld\altaffilmark{7}}}
\author{\mbox{C. Qiu\altaffilmark{2}}}
\author{\mbox{S. Liu\altaffilmark{2}}}

\affil{\altaffilmark{1}School of Astronomy and Space Science, University of Chinese Academy of Sciences, Beijing 100049, China}

\affil{\altaffilmark{2}School of Physics and Astronomy, Sun Yat-sen University, Zhuhai Campus, 2 Daxue Road, Xiangzhou District, Zhuhai, P. R. China {\it \rm napolitano@mail.sysu.edu.cn}}

\affil{\altaffilmark{3}National Astronomical Observatories, Chinese Academy of Sciences, 20A Datun Road, Chaoyang District, Beijing 100012, China}

\affil{\altaffilmark{4}CSST Science Center for Guangdong-Hong Kong-Macau Great Bay Area, Zhuhai, China, 519082}

\affil{\altaffilmark{5}INAF -- Osservatorio Astronomico di Capodimonte, Salita Moiariello 16, 80131 - Napoli, Italy }

\affil{\altaffilmark{6}Carmel College, Mala, Thrissur -680732, Kerala, India}
\affil{\altaffilmark{7}Leiden Observatory, Leiden University, PO Box 9513, NL-2300 RA Leiden, the Netherlands}

\begin{abstract}
Next generation large sky surveys will observe up to billions of galaxies for which basic structural parameters are needed to study their evolution. This is a challenging task that, for ground-based observations, is complicated by seeing limited point-spread function (PSF). To perform a fast and accurate analysis of galaxy surface brightness, we have developed a family of supervised Convolutional Neural Networks (CNN) to derive Sersic profile parameters of galaxies. This work presents the first two Galaxy Light profile convolutional neural Networks (GaLNets) of this family. The first one is trained using galaxy images only (GaLNet-1), and the second is trained with both galaxy images and the local PSF (GaLNet-2). We have compared the results from the GaLNets with structural parameters (total magnitude, effective radius, Sersic index etc.) derived on a set of galaxies from the Kilo-Degree Survey (KiDS) by 2DPHOT, as a representative of ``standard'' PSF convolved Sersic fitting tools. The comparison shows that GaLNet-2 can reach an accuracy as high as 2DPHOT, while GaLNet-1 performs worse because it misses the information on the local PSF. Both GaLNets are three orders of magnitude faster than standard methods in terms of computational speed. This first application of CNNs to ground-based galaxy surface photometry shows that they are promising tools to perform parametric analyses of very large galaxy samples, like the ones expected from Vera Rubin/LSST surveys. However, the GaLNets can be easily modified for space observations from Euclid and the China Space Station Telescope.
\end{abstract}

\section{Introduction}
\label{sec:intro}
Galaxy structural parameters, such as the total magnitude $mag$, the effective radius $R_{\rm eff}$ and S{\'e}rsic $n$-index, are basic measurements for
many empirical scaling relations. They are used for the size--mass relation (\citealt{Shen+03}; Baldry et al. 2012, Lange et al. 2016, \citealt{Roy+18}), the $\mu_e–R_e$ relation \citep{Kormendy77_II, Capaccioli+92a}, the fundamental plane \citep{Dressler+87}, the Faber--Jackson relation \citep{FJ76}, etc.

The dependence of these relations on galaxy properties, such as the environment where galaxies reside and their evolution with redshift, is a fundamental test of the galaxy formation and evolution scenario,
as they provide evidence of the growth in size and mass of these systems and the processes behind them. For instance, within the hierarchical model in the $\Lambda$--cold dark matter cosmology, most of present-day massive, spheroidal galaxies are expected to have grown in size and mass via a two--phase assembly.
A first phase, at redshift $1<z<2$, when galaxies turn to red compact systems (likely the progenitors of the cores of today's ellipticals, e.g. \citealt{Wellons+15_lower_z, Wellons+15_z2}) and a second phase (e.g., Oser et al., 2012), characterized by mergers and gas inflows, leading to rejuvenated star formation and dramatic size evolution from $z\sim$1 to today (\citealt{Trujillo+07}; \citealt{Buitrago+08}).
On the other hand, less luminous/massive spheroids, i.e. those having magnitude fainter than the knee of the luminosity function, $m_*$, and a stellar mass smaller than $M_*\lsim10^{10.5}M_\sun$, are predicted to have a less dramatic size evolution (e.g. Furlong et al. 2017). Similarly,  star--forming galaxies also show a milder evolution than spheroids (see e.g. van der Vel 2014; \citealt{Roy+18}). 

Overall, hydrodynamical simulations seem to show that the different growth between passive spheroids and active disc-dominated systems is mainly due to the fraction of accreted mass, which is higher in the former, especially in the higher mass bins (e.g. \citealt{Oser+10}; Furlong et al. 2017). 

To fully test these predictions, 
it is crucial to 
measure structural parameters \rui{for} large statistical samples of galaxies over a wider range of redshifts and stellar masses.

Next generation large sky surveys like Vera Rubin/LSST (\citealt{Izevic+19_LSST}), Euclid mission (\citealt{Laureijs+11_Euclid}), the Chinese Space Station (CSST, \citealt{Zhan+18_csst}), will eventually provide this unique opportunity.  
For instance, with the Vera Rubin/LSST full depth  (5$\sigma$ point source $r_{\rm AB}\sim27.5$ coadded), we will be able to identify $m_*+2$ galaxies in clusters and field at $z\sim1.5$, and reach a signal-to-noise ratio high enough to study the galaxy structural parameters of $m_*$ galaxies at $z\sim1.2$ in $ugrizy$ optical bands (\citealt{Robertson+17_LSST}). Similarly, CSST and Euclid will allow accurate surface brightness measurements for $\sim 2\times 10^8$ galaxies at $1\leq z <3$ in UV/optical and near-infrared, respectively (see e.g. \citealt{Laureijs+11_Euclid}). 

Hence, in the oncoming years, we will analyze the surface photometry of billions of galaxies in many optical and near-infrared bands, which is a major technological and methodological challenge. If, on the one hand, these measurements need to be performed in a reasonable timescale, on the other hand, 
the galaxy complexity to account for in the modelling 
(e.g. multi-component vs. single component systems)
will make the analysis process computationally challenging. 

So far, most of the analyses of galaxy structural parameters over large datasets have been mainly based on one-component galaxy models, e.g. using of the S{\'e}rsic (1968)
profile. This has been successfully used to measure the main structural parameters of galaxies in different datasets, e.g. Sloan Digital Sky Survey (SDSS, e.g., \citealt{Shen+03}, \citealt{HB09_curv}), GAlaxy Mass Assembly (GAMA, e.g., \citealt{Baldry2012}, \citealt{Lange2015}), Kilo Degree Survey (KiDS, e.g. \citealt{Roy+18}), Dark Energy Survey (DES, e.g., \citealt{Tarsitano+DES+sersic}), Hyper Suprime-Cam Subaru Strategic Program (HSC, e.g. \citealt{Kawinwanichakij2021}), and the Cosmic Assembly Near-infrared Deep Extragalactic Legacy Survey (CANDELS, e.g., \citealt{Shibuya2015ApJS..219...15S})

Indeed, with only three free structural parameters (the total magnitude $mag$, the effective radius $R_{\rm eff}$, and the S{\'e}rsic index $n$), together with some other shape/position parameters (the axis ratio $q$, the position angle $PA$ and the galaxy center coordinates $x_{\rm cen}, y_{\rm cen}$), 
this model has a sufficient generality to match most of the light distribution of galaxies of different morphology (i.e. early vs. late-type galaxies). Also, 
it has been successfully used to reproduce the surface brightness distribution of galaxies from the inner sub-arcsec scales to relatively larger radii, even for ground-based observations (see e.g. \citealt{Tortora+18_UCMGs}). 
Obviously, to fully catch the physics of galaxy evolution, it is mandatory to move toward more complex models considering multi-component profiles (see e.g. \citealt{Dimauro+18}). 
{However, besides the complication of the higher degeneracy among the parameters, {which is worsened by the low spatial resolution of ground-based observations,} these models do add further computational complexity}.

Currently, traditional galaxy fitting codes, like GALFIT (\citealt{Peng2002+Galfit}), Gim2d (\citealt{Simard2002+Gim2d}), and 2DPHOT (\citealt{LaBarbera_08_2DPHOT}),
are either too slow
or need too much manual intervention. 
Hence, they could satisfy the required speed demand when suitably parallelized and/or using next-generation multi- CPU machines. 
Although ``hybrid codes", like GALAPAGOS (\citealt{Barden2012+GALAPAGOS}) or 
PyMorph (\citealt{Vikram2010+PYMORPH}),
are trying to overcome these technical difficulties, the computational time 
per galaxy is still rather high and strongly dependent on the number of model parameters. Furthermore, for most of these codes, the accuracy and success rate is also strongly dependent on the initial conditions(see e.g. \citealt{Yoon+11}). 
Hence, faster, automatic, and accurate methods are highly demanded for next generation sky surveys.

Machine Learning (ML) tools 
possibly provide the most viable solution to this need.
In the last years, they have been successfully applied to many fields of astronomy
to solve typical classification or regression tasks 
with fast speed and high accuracy. For instance, they have been used to search for strong gravitational lenses either from images (\citealt{Petrillo+17_CNN, Petrillo+19_CNN,Petrillo+19_LinKS,Jacobs+19, Canameras+20_Holismokes-II,Li+20_KiDS,Li+2021+DR5lens}) or from spectra (\citealt{Li+19}); for automatic extraction of spectral features (Wang, Guo \& Luo 2017); for unsupervised feature-learning for galaxy SEDs (Frontera-Pons et al. 2017); and for photometric redshifts (\citealt{Bilicki2018+MLphotz}).

%
Among the ML tools above, Convolutional Neural Networks (CNNs), in particular, allow us to
optimally handle problems related to image processing and feature recognition that are suitable for galaxy morphology and surface brightness analysis. Indeed, galaxy classification 
has been one of the fields where CNNs have produced early promising results. \cite{Dieleman2015MNRAS} developed a CNN model to classify the galaxies that have been labelled by human inspection in the Galaxy Zoo project. The CNN reached 99\% accuracy, while having a much faster speed than human classification. Similarly, \cite{Tarsitano2021arXiv} developed a CNN-based approach to classify the galaxies in DES, and achieved 86\% accuracy for early-type galaxies and 93\% for late-type galaxies.

CNNs have also been successfully applied to 
a variety of galaxy photometry related studies, such as non-parametric light profile extraction (e.g., \citealt{Smith2021MNRAS, Stone2021MNRAS}), source deblending (e.g., \citealt{Boucaud2020MNRAS}), and galaxy stellar population analysis (e.g., \citealt{Buck2021arXiv}).

\cite{Tuccillo+18} for the first time applied CNNs to two-dimensional light profile galaxy fitting on 
HST/CANDELS data. Their code (DeepLeGATo) manages to fit galaxies with 
a single S{\' e}rsic profile, 
closely reproducing the results obtained with GALFIT, but with much shorter computational time and in a fully automatic way. 
To achieve the best accuracy, though, DeepLeGATo required some domain adaptation\footnote{{Domain adaptation is a special technique for transferring learning from a given domain encoded in a training sample into another domain characterising the predictive sample. This is a technique used to solve the discrepancies between the predictions and the true values that can come from a mismatch between the sample features used  to train a CNN and the real properties of the predictive sample (see \citealt{Csurka2017_domain_adaptation}). For instance, in the specific case of the galaxy structural parameters, the simulated galaxies can have no companion systems, while the real ones do (see \citealt{Tuccillo+18}.}}, 
because the initial simulated training set was not ``realistic" enough. 
Furthermore, being developed for space observations, DeepLeGATo was not designed to 
account for the seeing limited point-spread-function (PSF).


The impact of the local PSF has been investigated previously in \cite{Umayahara2020SPIE11452E..23U}, who showed that adding the information of PSF to the galaxy images in deep learning tools would improve 
the prediction accuracy for the effective radius in ground-based observations.
However, this early attempt did not test the full constraint of a S\'ersic model, including all other model and geometrical parameters like the $n-$index, the total magnitude, the axis ratio, and the PA. 
Hence, to our knowledge, there are currently no works that have 
produced science-ready ML tools to infer galaxy surface photometry, fully accounting for the PSF effect on ground-based observations.
For this reason, we have started a project to develop ML tools to analyse the 2D surface brightness of galaxies using primarily ground-based observations. The main aim is to produce a family of fast and accurate GAlaxy Light profile convolutional neural NETworks (GaLNets, in short) to perform single and multi S{\' e}rsic profile models of galaxies on different datasets. This will allow us to test the applicability of the GaLNets on these data for future ground-based surveys (e.g. Vera Rubin/LSST), but it will eventually be extended to space observations (e.g. Euclid and CSST) with other dedicated GaLNets.

In this first paper, we develop two GaLNets to fit single \cite{Sersic68} profile parameters to galaxies from public ground-based data of the Kilo Degree Survey 
(KiDS: \citealt{deJong+15_KiDS_paperI}; \citealt{deJong+17_KiDS_DR3}; \citealt{Kuijken+19_KiDS-DR4}), as a prototype of high quality ground-based imaging dataset, but this can be easily generalized to other imaging datasets. 
To train the two CNNs, we simulated realistic KiDS mock galaxies by 
fully taking into account the observed seeing to produce PSF-convolved 2D galaxy models that we add to randomly selected ``noise cutouts''. These mock observations are similar to what we have implemented in the strong lensing classifiers (see \citealt{Li+20_KiDS, Li+2021+DR5lens}), where we have produced realistic color images of gravitational arcs and multiple images from lensed sources. 
The two GaLNets differ by the ``features'' that they use in the training phase: the first one (GaLNet-1), is fed with only galaxy images as input, while the second one (GaLNet-2), is fed with both galaxy images and the ``local'' PSFs\footnote{\nic{With ``local'' PSFs, we refer to the fact that the network is trained with PSF models that take into account the characteristic spatial variation of the PSF across the observed images.}}. 
The reason for introducing GaLNet-1, which is expected to perform generally worse than GaLNet-2 because of lacking the PSF information, is to quantify the impact of the PSF on the accuracy of the results and possibly find room for improvements of the PSF model to adopt.

This paper is meant to provide the first analysis of the performance of this novel approach. For this reason, we use only $r$-band observations, as these are the best image-quality data in KiDS (see e.g. \citealt{Kuijken+19_KiDS-DR4}).  
We will also use a limited sample of $\sim 25\,000$ galaxies for which we have previously derived independent structural parameters (i.e. total magnitude, effective radii, S\'ersic index, axis ratio, etc.,  see \citealt{Roy+18}) using 2DPHOT, a PSF-convolved S{\'e}rsic fitting tool based on $\chi^2$ minimization (\citealt{LaBarbera_08_2DPHOT}). In order to avoid systematics due to the selection and modelling of the local PSF, we will adopt the PSF models produced by 2DPHOT for the galaxy sample (see Roy et al. 2018 for details) as input for the GaLNet-2.

In future papers, we will apply the GaLNets to the full multi-band KiDS 
dataset to study the structural properties of galaxies as a function of galaxy mass and redshift.
In future papers of this series, we will also expand the GaLNets model capabilities to fit multi-component systems and test their application to future survey facilities like Vera Rubin/LSST, Euclid and CSST. 

This work is organized as follows. In Sect. 2, we describe how to build the training and testing sample and describe the CNN architectures. In Sect. 3, we test our CNNs on simulated and real data. In Sects. 4 and 5, we make some discussions and summarize our main conclusions.


\begin{figure*}[htbp]
\centerline{\includegraphics[width=15.5cm]{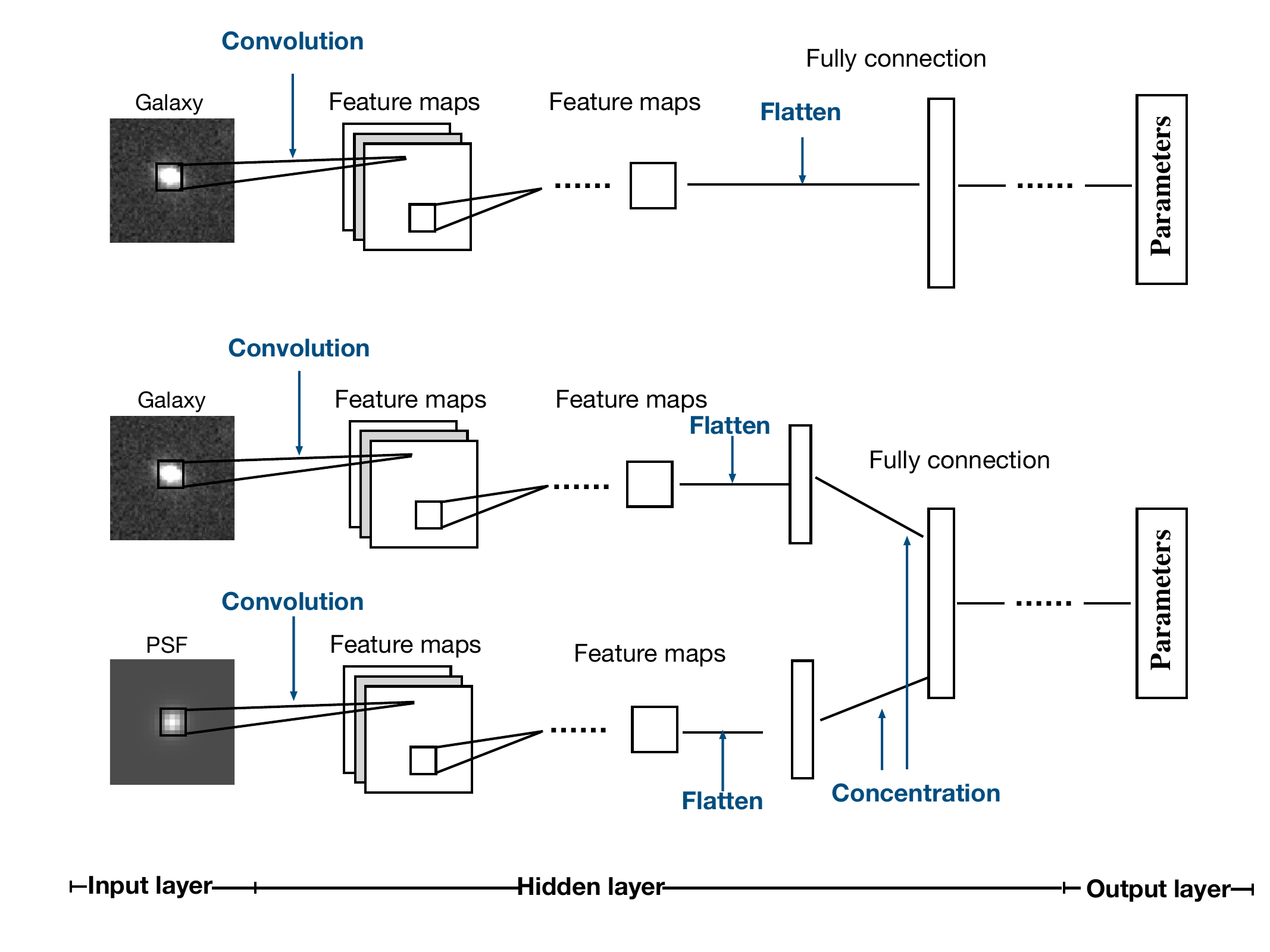}}
\caption{The GaLNet models used in this work. Up: structure of GaLNet-1, with only galaxy images as inputs. bottom: structure of GaLNet-2, fed by both galaxies images and the corresponding ``local'' PSFs from 2DPHOT.}
\label{fig:CNN_sample}
\vspace{0.5cm}
\end{figure*}

\section{The CNN method}
CNNs are a special category of Neural Networks made of layers of artificial neurons called nodes. The nodes contain the convolution part where specific activation functions calculate the weighted sum of the inputs and return an activation map. Convolution is able to save the important information contained in the ``features'' (e.g. a pattern or a color in an image) by simultaneously reducing the size of the elements carrying such relevant information. This produces a massive saving of computational time in high-resolution images, hence making CNNs particularly suitable for both classification and regression tasks related to image analysis.

CNNs are trained over specific samples containing the features to be used to make predictions on a series of ``targets'' (i.e. a pattern, a property, one or more parameters), over a sample of objects that represent the ``predictive data''. The crucial ingredients in this ``training'' phase are 1) the perfect correspondence between the sample used for the training ({\it training set}) and the sample for which we need to estimate the target parameters ({\it predictive sample}) and 2) the size of the {\it training set}, which needs to encompass the full target parameter space. These are often not available, as the ground truth values are not given for real objects one wants to analyze; and even if these latter are provided, they are limited to small statistics from other standard methods. A classic example for this latter case is strong gravitational lensing (see e.g. \citealt{Petrillo+17_CNN}; \citealt{Jacobs+19}; \citealt{Li+20_KiDS}), for which only a few hundred confirmed cases are available (e.g., SLACS, \citealt{Bolton2008+SLACS}; BELLS, \citealt{Brownstein2012+bells}; BELLSGALLERY, \citealt{Shu2016+bellsgallery}).

In this paper, we tackle a classic regression task where
the inputs are images of individual galaxies or pairs of images of galaxies and the corresponding PSFs, and the outputs are 7 parameters of the best S{\'e}rsic profiles describing the 2D galaxy light distribution.
This is defined as
\rui{
\begin{equation}
I(R)=I_e \exp \{-b_n[(\frac{\sqrt{(x-x_{\rm 0})^2+(y-y_{\rm 0})^2/q^2}}{R_{\rm eff}})^\frac{1}{n}-1]\},
\label{eq:sersic}
\end{equation}
}
where the $R_{\rm eff}$ is the effective radius, $I_e$ is the surface brightness at the effective radius, and ($x_{\rm 0}, y_{\rm 0}$) is the position of the galaxy center. The position angle $PA$ gives the angle between the ($x,~y$) plane and the sky coordinate frame, which is defined by assuming to be positive from North to East. Finally, $q$ is the axis ratio, and $n$ is the S{\'e}rsic index.
This latter is known to be a proxy of the galaxy morphology as early-type galaxies generally have $n\gtrsim2.5$ while late-type galaxies have $n\lesssim2.5$ (e.g. \citealt{Trujillo+07}). In Eq. \ref{eq:sersic}, for $b_n$ 
we use the expression provided by Ciotti \& Bertin (1999): 
\begin{equation}
	b_n \approx 
	\begin{cases}
	    2n-1/3+4/(405n),\ \ \   n\geqslant 0.36\\
	    0.01945-0.8902n+10.95n^2,\ \ \   n<0.36.
    \end{cases}
\end{equation}
The total (apparent) magnitude, $mag$, is defined by
\begin{equation}\label{mag}
    \centering
    mag = -2.5 \log (F_{tot})+zpt
\end{equation}
where $zpt$ is the zero point of the photometry. $F_{tot}$ is the total flux of the galaxies, definded as
\rui{
\begin{equation}\label{mag}
    F_{tot}=2\pi R_{eff}^2 I_e e^{b_n} n b_n^{-2n} \Gamma (2n) q,
\end{equation}
}
in which $\Gamma$ indicates the standard $\Gamma-$function. In summary, the fully 2D galaxy model contains 7 free parameters: $x_{\rm 0}, y_{\rm 0}$, $mag$, $R_{\rm eff}$, $q$, $PA$ and $n$.



The observed surface brightness profiles of galaxies result from the convolution
of the intrinsic (real) galaxy fluxes and the PSFs. Equally, the modelled 2D profile is given by the convolution of the 2D S\'ersic profile and the PSF model. This can be computed by the equation:

\begin{equation}\label{Eq:fitting}
    \centering
    M(BG,\{p_k\}) = BG + I(\{p_k\}) ~{\rm o}~ S
\end{equation}
where $I$ is the galaxy surface brightness distribution from Eq. \ref{eq:sersic}, $\{p_k\}$ are the model parameters, $S$ is the PSF model; $BG$ is the
value of the local background, and the symbol ``o" denotes convolution. In particular, for the PSF we adopt two Moffat profiles (see
\citealt{LaBarbera_08_2DPHOT}).

\subsection{The CNN architecture}
\label{sec:CNN}
Here below, we will describe the CNN method behind the GaLNets, including the customization of the CNNs architecture, the building of the training data, and the training of the CNN models.

CNN, which mimic the biological perception mechanism, is one of the representative deep learning algorithms. Depending on whether labels are provided for the training data, this learning process can be supervised or unsupervised. Supervised learning can infer a function from the labelled training data. These data are composed of input objects (usually vectors) and desired output values (labelled targets). 

The simple idea behind the GaLNets is that one can input the galaxy images into the CNNs and requires them to output the S{\' e}rsic parameters, having the CNN learn how labelled profiles
would look like in images with the same noise/background structure and seeing of real observed galaxies. However, especially for ground-based data, besides the intrinsic degeneracies among the S{\' e}rsic parameters in Eq. \ref{eq:sersic} (see e.g. \citealt{Trujillo2001MNRAS}), 
there are some further degeneracies among these latter and the PSF (see e.g. \citealt{Trujillo+01_II_PSF}).
As a very intuitive example, one can think of the intrinsic galaxy axis ratio. The PSF is, by definition, rounder than a flattened galaxy, hence the effect of the atmosphere on the 2D light distribution (i.e. a 2D convolution) produces as a net effect a rounder galaxy. The higher the seeing in arcsecs (which is measured by the full-width-at-half-maximum, FWHM),
the rounder the observed galaxy is.  
Without the knowledge of the PSFs, the prediction from the CNN could be biased toward lower axis ratios (i.e. rounder shapes). 

In this work, we want to explore 1) how does a CNN perform if only galaxy images labelled with the corresponding true intrinsic parameters are used in the training phase, and 2) how is a CNN improved if, together with the galaxy image, the corresponding 2D model of the local PSF is also inputted. 

To 
check these two schemes, we build two CNN models, GaLNet-1 and GaLNet-2, respectively. 
These are schematized 
in Fig. \ref{fig:CNN_sample} 
where we use, in particular, a slightly modified VGGNet \citep{Simonyan2014+VGGnet}. 
The overall structure is common to other CNNs, and consists of three parts: the input layer, the hidden layer, and the output layer. 
The heart of the CNN is the hidden layer, which includes: 1) the convolutional layer, 2) the pooling layer, and 3) the fully connected layer. The convolutional layer, containing  multiple convolution kernels composed of weights and bias, is used to extract features from the input data. These ``feature maps'' will be passed to the pooling layer for information filtering to reduce their intrinsic size. The pooling layer contains a pre-set pooling function used to replace the feature map in a given region with a single point (flattening). The fully connected layer is located at the end of the hidden layer of networks to perform a non-linear combination of the extracted features. 
In particular, it combines the low-level learned features into high-level features and passes them to the output layer.

For GaLNet-1, with only galaxy images as input,
the hidden layers are made of
4 convolutional blocks, and in each block there are two weight layers, followed by an averaging $2\times 2$ pooling layer. These convolutional blocks return flatted feature maps and the feature maps are then fed 
to 3 fully connection layers 
to combine the low-level features to higher-level ones. In this phase, the CNN makes the decision about the 
7 S{\' e}rsic parameters to output through the output layer. 

For GaLNet-2, including also the 2D PSF model images as input (see Sect. \ref{sec:training_data} for details), we designed a two-path architecture: the main path is used to process the images, and the secondary path is used to process the PSFs. The main path is the same as that in GaLNet-1.  The secondary path has only one convolutional block composed of two weight layers. After the flattening, the two paths are concentrated and then followed by 3 fully connected layers. Here the CNNs make the decision about the 7 S{\' e}rsic parameters after having learned the PSF morphology and output the result through the output layer.

\subsection{Training and testing data}
\label{sec:training_data}
\begin{figure*}[htbp]
\centerline{\includegraphics[width=18cm]{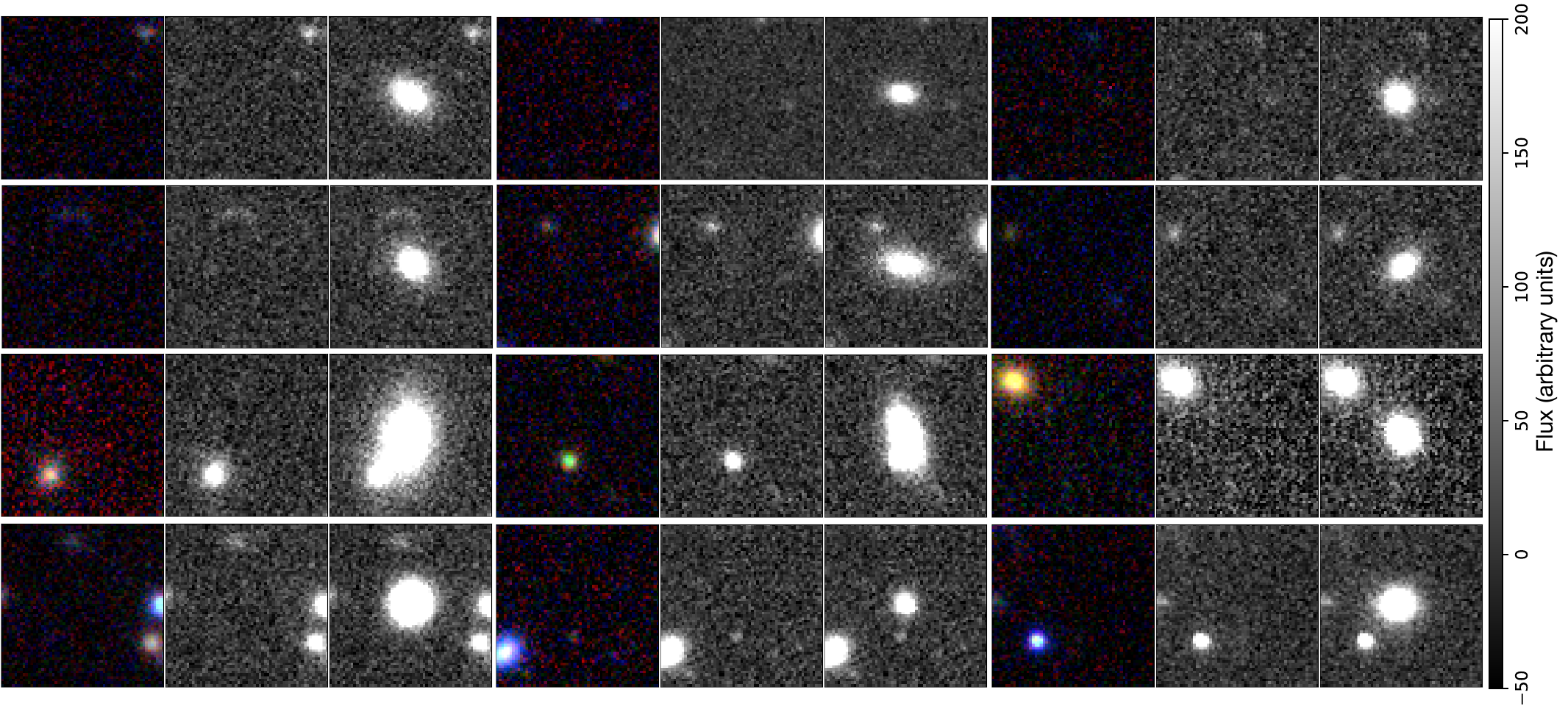}}
\caption{Examples of simulated galaxies for training the GaLNets. For each three-panel image we have the $gri$ color combined (left panel), the r-band image (central panel) of the background cutouts, and the simulated galaxy added in the cutout center (right panel), according to its labelled S\'ersic parameters. We can see the variety of real ``contaminants'' in the background cutouts, including bluer and redder stars and galaxies of different sizes and colors.}
\label{fig:simulation}
\vspace{0.5cm}
\end{figure*}

The training set is built by adding 2-dimensional, PSF convolved, simulated S{\'e}rsic profiles to randomly selected $r-$band cutouts from public KiDS data, representing the galaxy ``background''. These random background images are not 
empty but can contain other sources 
to simulate realistic environment situations (see Fig. \ref{fig:simulation} and description below). 

The mock galaxy images constituting
the training data are obtained with the following procedure:
\begin{enumerate}
\item {{\it Background images.} We} randomly obtained 20\,000 cutouts with a size of $61\times61$ pixels from the $r-$band observations in KiDS. We visually make sure that 1) these cutouts are not located in the areas with saturated stars or stellar diffraction spikes or reflection haloes; 2) there are no bright galaxies or stars in the central region. These cutouts represent the ``background sample" as introduced above;
\item {{\it PSFs.} We} collect 25\,000 PSFs from previous 2DPHOT runs (see e.g. \citealt{Roy+18}), where
2D local PSF models paired to each galaxy 
are obtained assuming two-Moffat profiles (see \citealt{LaBarbera_08_2DPHOT}). We obtain $25\times25$ pixels images of the modelled PSF that make up the ``PSF sample". The mean  FWHM of the PSFs is 0.70$''$, and the standard deviation is 0.11$''$, similar to the one of the full DR4 in \cite{Kuijken+19_KiDS-DR4};
\item {{\it Mock galaxies.} We} simulate the S{\'e}rsic profiles with a range for the 
7 parameters 
as shown in Tab \ref{tab:parameters}. The distribution of $x_{\rm cen}$, $y_{\rm cen}$ and $\log R_{\rm eff}$ are Gaussian, $mag$ is exponential, $n$ is F, while $q$ and $PA$ are uniform.
These distributions are chosen to mimic observed parameters distributions in galaxies (see e.g. \citealt{Roy+18}).
\item {\it PSF convolution and final stamps.} We convolve each S{\'e}rsic profile with a randomly selected two-dimensional PSF from the PSF sample, then a random Poisson noise is added.
The resulting 2D profile is finally summed to a random ``background" cutout (from step 1),
to obtain fairly realistic galaxy images. \nic{If the signal-to-noise ratio (SNR), as computed over a central circular area of radius 1.5$''$,  is larger than 50,} this galaxy is kept as training data, otherwise discarded.
\end{enumerate}
Adopting an $SNR>50$ for the training sample is made for uniformity with typical choices made for real galaxies to ensure reasonable accuracy on the fitted parameters (see e.g. \citealt{LaBarbera_deCarvalho09}, \citealt{Roy+18}).
For this first analysis, we do not plan to test the impact of the SNR on GaLNets predictions, which we will leave for future analyses (see also Sect. \ref{sec:future})
 
\begin{deluxetable}{llcl}
\centering
\tablecaption{parameters for simulating the training sample \label{tab:parameters}}
\tablewidth{0pt}
\tablehead{Parameter & Range &units& Distribution}
$x_{\rm cen}$&-0.4 -- 0.4&arcsec& normal\\
&  && ($\mu=0.0$, $\sigma=0.15$)\\
$y_{\rm cen}$&-0.4 -- 0.4&arcsec& normal\\
&  && ($\mu=0.0$, $\sigma=0.15$)\\
$mag$&17 -- 22& -- & exponential\\
&  && ($s=1$)\\
$R_{\rm eff}$ & 0.2 -- 4 &arcsec& normal ($\log R_{\rm eff})$\\
&  && ($\mu=-0.1$, $\sigma=0.4$)\\
$q$ & 0.2 -- 1.0 &--& uniform\\
$pa$ & 0.0 -- 180.0 &degree& uniform\\
$n$ & 0.1 -- 8.0 &--& F\\
&  && ($n=30$ $d=5$)
\tablecomments{Range and distribution of parameter values used to simulate the galaxies. $\mu$ and $\sigma$ are the mean value and standard Deviation of a normal distribution. $s$ is the scale of the exponential distribution.  $n$ is the degrees of freedom in numerator and d is the degrees of freedom in denominator.}
\label{tab:parameters}
\end{deluxetable}

We also remark that the use of realistic distributions of parameters as ``prior'' for the training sample, in step 3, is based on a series of tests made on different distributions (see Appendix \ref{sec:appendix}). 
To summarise, we find that the unrealistic distributions of $n$-index will produce some systematic features at $n>4$, while for other parameters, the use of unrealistic priors will affect very little the final predictions.
This implies that, although we have paid attention to reproducing realistic enough ground truth parameter distributions for the $n-$index, we can not exclude that there are still some systematics with respect to the intrinsic parameters. For this reason, we will explore the use of Bayesian Neural Network for future developments (see e.g. \citealt{Gal2015}, \citealt{Wagner-Carena2021}). This will be a natural choice to include error estimates to the individual predicted parameters, that are not available with the current GaLNets. 
For the time being, as shown in Appendix \ref{sec:appendix}, we stress that for variations as large as 30\% in the parameter distributions, the GaLNets do not suffer significant biases, except for large values of the S{\'e}rsic index ($n>4$).

Finally, following steps 1--4 above, we simulate 220\,000 mock galaxies.
We split this sample into the {\it training data}, to which we assign 200\,000 mock galaxies, and the {\it test data}, made of the residual 20\,000 simulated galaxies. 
In Fig. \ref{fig:simulation} we show some mock galaxies and their original background images. There are clearly isolated cases with empty backgrounds, and  cases with close systems of different sizes, shapes and colors. Here we remark that, according to the procedures above, the PSFs of the simulated galaxies do not match the ones of the 
background cutouts, but are randomly selected from the ``PSF sample". 
However, we do not expect this to affect the final results, as the CNNs focus only on the model of the central galaxy. However, this is a refinement that we will implement in future analyses when the GaLNets will implement a self-made PSF model.

On the other hand, a parameter that can impact the final CNN predictions is the size of the cutouts. We have taken 61 pixels ($\sim12''$) by side because this provides an area large enough to sample most of the light profile of galaxies with $R_{\rm eff}\sim1''$ (corresponding to $\sim$ 5 pixels). This latter represents the majority of the galaxies observed in typical ground-based observations, like KiDS (see e.g. Fig. 9 of \citealt{Roy+18}). However, for larger galaxies, the fraction of the total light enclosed in the cutout can be rather different. For instance, by 
integrating Eq. \ref{eq:sersic} (without seeing convolution ) over the sky plane and considering the galaxy centered in the cutout with $12''$ by side, the
fraction ($f$) of total projected luminosity within $12''$ diameter is:
\begin{enumerate}
\item[$\bullet$]{$f=99.9\%$ for $R_{\rm eff}=1''$ and $n=1$,}
\item[$\bullet$]{$f=91\%$ for $R_{\rm eff}=1''$ and $n=4$,}
\item[$\bullet$]{$f=86\%$ for $R_{\rm eff}=1''$ and $n=6$,}
\item[$\bullet$]{$f=61\%$ for $R_{\rm eff}=4''$ and $n=4$.}
\end{enumerate}

Hence, only for galaxies with small $R_{\rm eff}$ ($<1''$) and small $n$ ($<6$), most of the light is collected into a $12''$ wide cutouts. For larger $n$ even for $R_{\rm eff}\sim1''$ there is a significant fraction of the light emitted outside the cutouts. Finally, for larger $R_{\rm eff}$, only a small fraction of the whole galaxies is included in the cutout. 
Note that, using a relatively small cutout, might affect the parameter estimates for standard 2D fitting techniques (e.g. GALFIT \citealt{Peng2002+Galfit}; 2DPHOT \citealt{LaBarbera_08_2DPHOT}), because both the outer parts of the galaxy and the background are poorly sampled. However, investigating this issue is outside of the scope of the present paper.

For this reason, we have additionally tested 
larger cutout sizes, namely 101 and 151 pixels by side, corresponding to $\sim20''$ and $30''$ respectively, to check how the cutout sizes would affect their results. 
We have found that
the GaLNets generally give poorer results for small galaxies, while they do not show significant improvements for larger galaxies (i.e. $R_{\rm eff}>3''$), despite the background around the galaxies is better sampled. A reason for this result is that the CNN 
learns how to derive central parameters ($R_{\rm eff}$, $n$, $I_{\rm e}$, see Eq. \ref{eq:sersic}), which determine the gradient of the light profiles in the central regions. Eventually, the CNNs learn how to infer these parameters from the information they extract from the central steeper light gradient rather than the shallower slopes of the outermost radii. 
Hence, for the main analysis of this paper, we keep the 61 pixels cutouts as the best compromise for most of the systems we will analyze.


\begin{figure*}[htbp]
    \centering
    \subfigure {\includegraphics[width=18.0cm]{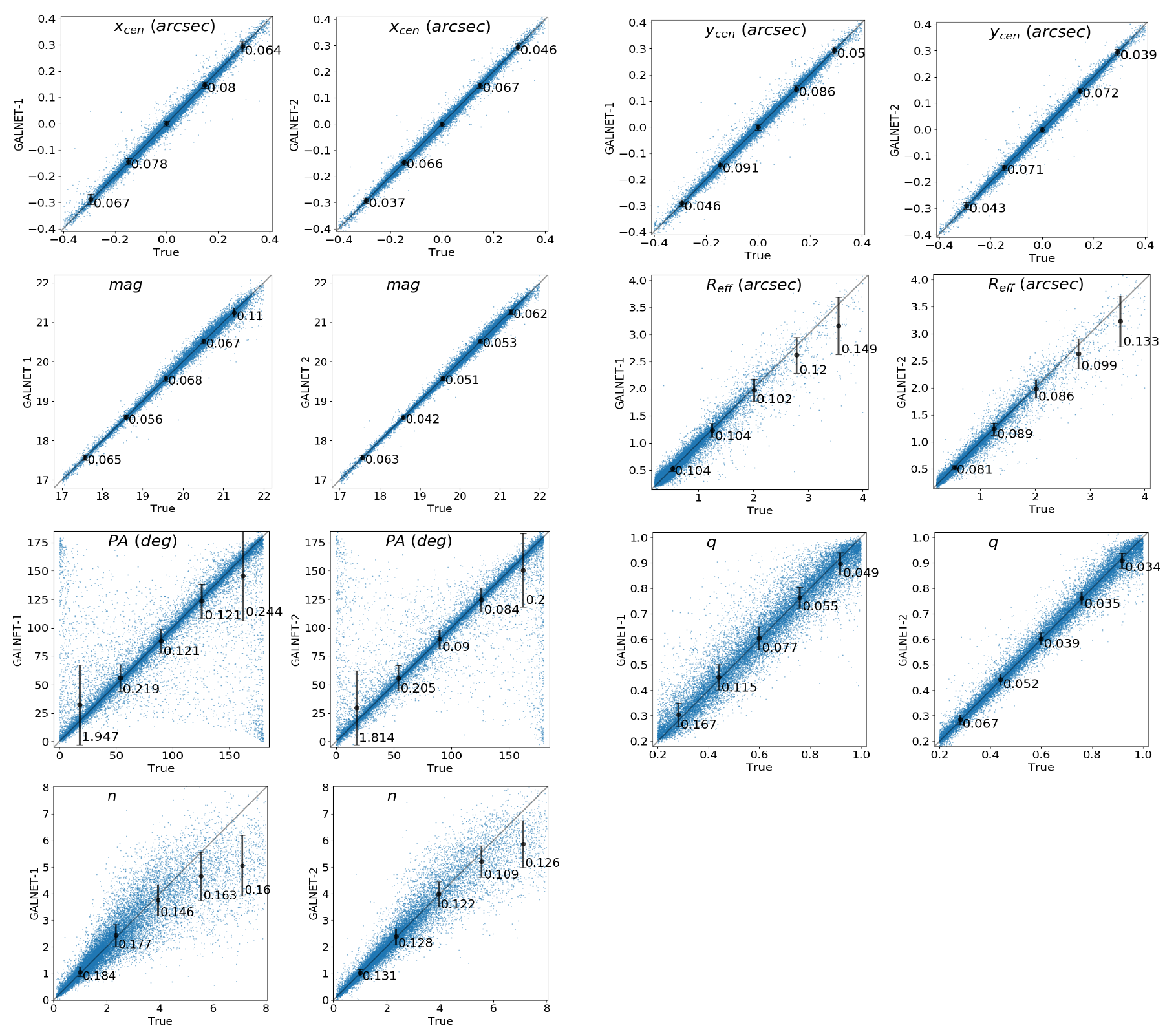}\vspace{3pt}}
    \vspace{0.2cm}
    \caption{ Comparison between the true value and the predicted value from the two GaLNets on simulated data. In each panel, the horizontal axis are the true values and the vertical axis are the predictions from the GaLNets. Error bars are the absolute mean errors in each bin, while labels report the absolute errors for $mag$ and relatives errors for others.}
    \label{fig:test_simulation}
    \vspace{0.2cm}
\end{figure*}

\begin{table*}[htbp]
\begin{center}
\caption{\label{tb:R2_simulation} Statistical properties of simulated data.}
\begin{tabular}{c c c c c c c c c}
\hline \hline
CNN model & $x_{\rm cen}$ & $y_{\rm cen}$ & $mag$ & $R_{\rm eff}$ &  $PA$ &$q$ &$n$ \\
\hline
\multicolumn{8}{c}{$R^2$}\\
\hline
GaLNet-1   & 0.9885  & 0.9882  & 0.9898  & 0.9206  & 0.6716  & 0.9124  & 0.7280 \\
GaLNet-2  & 0.9924  & 0.9915  & 0.9945   &0.9624   &0.7692.  & 0.9781  & 0.8905  \\
\hline
\multicolumn{8}{c}{fraction of outliers}\\
\hline
GaLNet-1 & 0.001& 0.001& 0.000& 0.028& 0.210& 0.008& 0.260 \\
GaLNet-2 & 0.000& 0.001& 0.000& 0.017& 0.165& 0.001& 0.128 \\
\hline
\multicolumn{8}{c}{NMAD}\\
\hline
GaLNet-1   &0.0100& 0.0099& 0.0028& 0.0420& 0.04840& 0.02678& 0.1094 \\
GaLNet-2  &  0.0090& 0.0090& 0.0022& 0.01990& 0.0344& 0.0139& 0.0655\\
\hline \hline
\end{tabular}
\end{center}
\textsc{Note.} ---Statistical properties of the predictions on the simulated testing data. From up to bottom we show the $R^2$, the fraction of outliers and the NMAD for the central position $x_{\rm cen}, y_{\rm cen}$, magnitudes $mag$, effective radius $R_{\rm eff}$, position angle $PA$, axis ratio $q$, and the S{\'e}rsic index $n$. The parameters of GaLNet-2 are always better than that of GaLNet-1.
\end{table*}

\subsection{Training the Network}
\label{sec:Training}
The input of the CNNs is the $61\times61$ pixels galaxy images for both GaLNet-1 and GaLNet-2 and the $25\times25$ pixels of PSF images for GaLNet-2 only. The outputs are the 7 parameters describing the light distribution of a S{\'e}rsic galaxy: $x_{\rm cen},~ ~y_{\rm cen}, ~mag, ~R_{\rm eff}, ~n, ~q, ~PA$).
We train the CNNs by minimizing the ``Huber" loss function (\citealt{Friedman99+huberloss}) with an ``Adam" optimizer (\citealt{Kingma2014+Adam}). 
The ``Huber" loss is defined as

\begin{equation}
	L_{\delta}(a) =
	\begin{cases}
	    \dfrac{1}{2}(a)^2,  \ \ \ |a|\leq\delta\\
	    \delta\cdot(|a|-\dfrac{1}{2} \delta),  \ \ \ {\rm otherwise}.
    \end{cases}
\end{equation}
in which $a=y_{\rm true}-y_{\rm pred}$, where $y_{true}$ is the real value for the simulations and $y_{\rm pred}$ is the predicted value by the CNNs. $\delta$ is a parameter that can be pre-setted. Given $\delta$ ($\delta$ was fixed to be $0.001$ in this work), the loss will be a square error when the prediction deviation $|a|$ is smaller than $\delta$, otherwise, the loss reduces to a linear function.
We have preferred the ``Huber" loss to the standard loss functions like the
Mean Square Error (MSE) and the Mean Absolute Error (MAE), because we have found the former to provide better accuracy and robust convergence. 
The reason to discard the MSE is that this latter 
gives higher weights to outliers, but at the expense of the prediction accuracy of other normal data points. Some of our targets are prone to outliers, see e.g. the $PA$ that can become rather random when $q$ is close to $1$ and degenerate between 0 and 180 deg when the ground truth is close to either of these two values, hence we want to avoid introducing artificial biases.
The MAE is less sensitive to outliers, but
it tends to give convergence problems as it does not properly weight the gradient in the loss function with the errors (for instance, it does not allow large gradients to converge in the presence of small errors), 
and the updated gradient in the training process is always the same. Therefore, even for a small loss value, the gradient is large, affecting the convergence speed of the training process, and sometimes the CNN even can not find the best fitting value.

In terms of computational time, using an NVIDIA RTX 2070 graphics processing unit (GPU), the training times for GaLNet-1 and GaLNet-2 have no significant differences, as both need about 2 hours to be trained with the 200\,000 galaxies sample. As we checked that the training time scales with the sample size (e.g. a 50\,000 sample takes 0.5 hours), then this means that we can train the CNN over a 2 million sample in less than one day with commercial GPUs.

\section{Testing the performances}
\label{sec:testing}
\subsection{Testing on simulated data}
\label{sec:testing_simulation}

After training the two GaLNets, we use the test sample (see Sect. \ref{sec:training_data}) to check their performances. In Fig. \ref{fig:test_simulation}, we compare the ground truth values (x-axis) of each parameter used to simulate the galaxies and the predicted values (y-axis) from GaLNet-1 (left panel) and GaLNet-2 (right panel).
To assess how strong the linear relationship is between the two variables in these plots, we adopt 3 diagnostics: 1) the R--squared ($R^2$), 2) the fraction of outliers and 3) the normalized median absolute de-viation (NMAD).

The $R^2$ is defined as
\begin{equation}
	R^2=1-\frac{\sum\limits_i({p}_i-t_i)^2}{\sum\limits_i(t_i-\bar{t})^2},
\end{equation}
where $t_i$ are the ground truth values, $p_i$ are the GaLNets' predicted values and $\bar{t}$ is the mean value of $t_i$. According to this definition $R^2$
is 0 for no correlation and 1 for the perfect correlation.  

The fraction of outliers is defined as the fraction of discrepant estimates larger than the 15\%, according to the following formula:
\begin{equation}
	|\Delta p|=\frac{|p_i-t_i|}{1+t_i}>0.15.
\end{equation}
This definition is 
usually 
adopted for outliers in photometric redshifts determination (see detail in \citealt{Amaro2021+photz}).
Finally, NMAD is defined as
\begin{equation}
	{\rm NMAD}=1.4826\times {\rm median} (|\Delta p- {\rm median }(\Delta p)|).
\end{equation}



The results obtained
for all these parameters are listed in Table \ref{tb:R2_simulation}.
From this Table and from Fig. \ref{fig:test_simulation}, 
we see that the predictions from GaLNet-1 and GaLNet-2 are generally good as the $R^2$ is close to 1. 
This is particularly true for $x_{\rm cen}, ~y_{\rm cen}, ~mag, ~R_{\rm eff}$ and $q$, with GaLNet-2 showing  $R^2$ systematically larger than GaLNet-1 (see below). 
The $R^2$ for position angle $PA$ and S{\' e}rsic index $n$ are slightly worse. For $PA$, this comes from a large scatter around $PA=0$ deg and $PA=180$ deg where the CNNs tend to concentrate the rounder systems and can be easier to swap the 0 deg with the 180 deg solutions.
For S{\' e}rsic index $n$, the poorer correlation is caused by the scatter at larger $n$ values. 
This mainly comes from two factors. First, the central slope of a S{\' e}rsic profile with a larger S{\' e}rsic index is very steep, which means the light decreases quickly within the central 1-2 pixels, so it is hard for the CNNs to fully ``measure'' such a steep variation, given the combination of the KiDS pixel size ($0.2''$) and the PSF ($\sim 0.6''-0.8''$ in the KiDS $r-$band).
Second, the outer part of such a large S{\' e}rsic--index profile is flat and is often hidden within the background noise, hence it can be hard for the CNNs to detect a significant light gradient that can help to guess about the $n$-values, as learned from the training sample. This effect is even more severe if the CNNs have no information about the PSF. This latter increases the degeneracies both at the low scales (where the seeing suppresses the peaks) and at large scales where the seeing tends to smooth the profiles and make a shallower slope (see e.g. \citealt{Tortora+18_UCMGs}). 
Encouragingly, the effective radius is a well constrained parameter from both GaLNets, despite the larger uncertainties on the $n$-index.
This can be quantified by the fraction of outliers and the NMAD as a proxy of the scatter, as reported in Table \ref{tab:parameters}. Here we see that the NMAD of $R_{\rm eff}$ is $0.07$ for GaLNet-1 and $0.03$ for GaLNet-2, comparable to the ones of the well constrained $mag$ and $q$. However, the fraction of outliers for $R_{\rm eff}$ (0.028 for GaLNet-1 and 0.017 for GaLNet-2) is larger than the ones of $mag$ (0.0 for both CNNs) and $q$ (0.008 and 0.001, respectively), although rather acceptable, especially for GaLNet-2. Finally, the $n-$index shows fractions of outliers of the order of $0.26$ and $0.13$, which prompts us to find strategies for improvement. This would also help to improve the result on $R_{\rm eff}$, discussed above. We remark, though, that the S{\' e}rsic index is traditionally a hard parameter to constrain from the ground, regardless of the tools adopted (see e.g. \citealt{Trujillo+01_II_PSF}), hence we consider the absence of systematics seen in Fig. \ref{fig:simulation} as an encouraging result of these first GaLNets.

\begin{figure*}[htbp]
\centerline{\includegraphics[width=18cm]{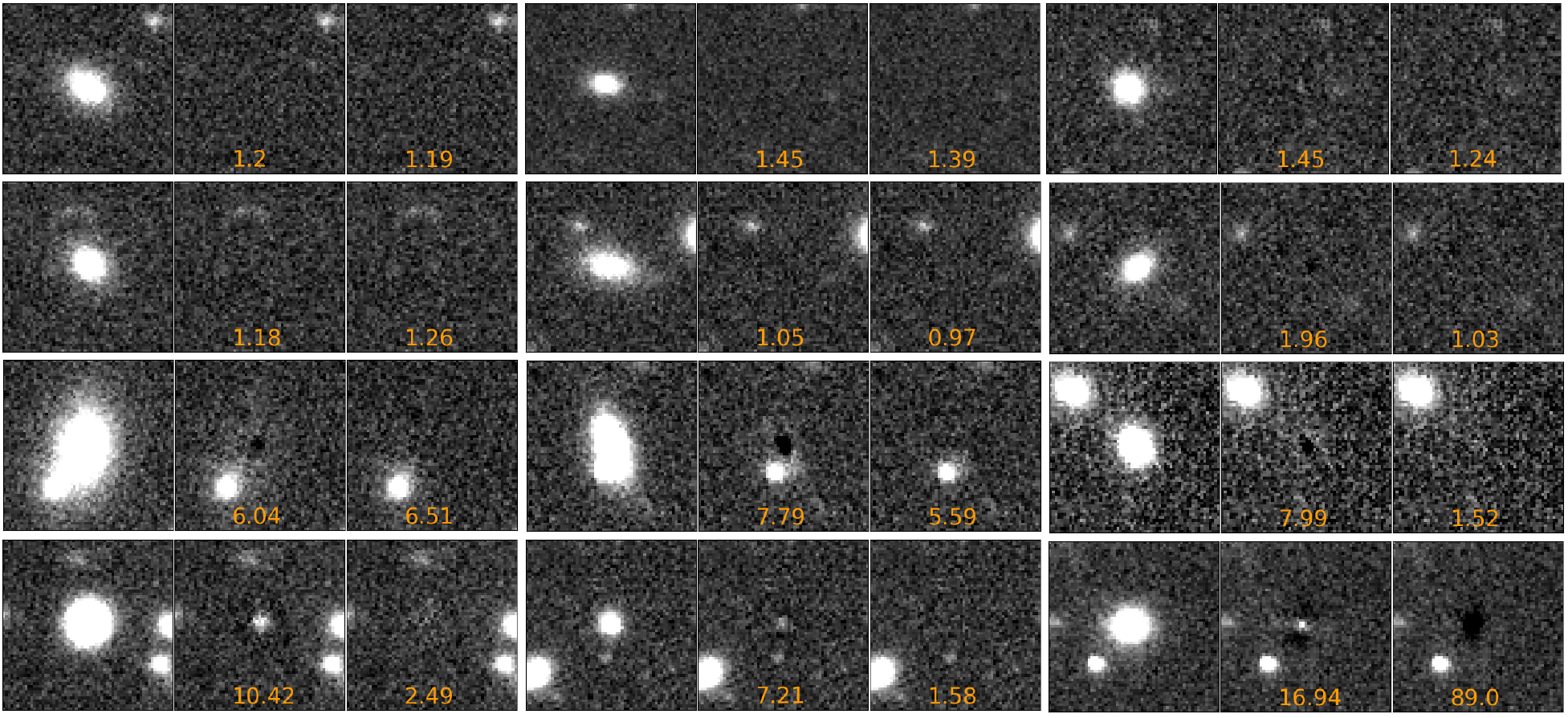}}
\caption{Residual maps, obtained from simulated "Galaxy" images (left panels), after subtracting the reconstructed S{\'e}ersic models using GaLNets-1 (middle-) and GaLNets-2 (right-panels). At the bottom of each residual image, we report the a posteriori reduced $\tilde{\chi}^2$ (see Sect. \ref{sec:testing_on_real} for definition).
}
\label{fig:residual_simulation}
\vspace{1.0cm}
\end{figure*}

Going into more details about the performances of the two GaLNets, 
the fact that GaLNet-2 always performs better than GaLNet-1
shows the importance of the information that the CNN can learn from the ``local'' PSF, especially for the galaxy shapes. Indeed, the improvement is more evident for the axis ratio $q$ and the S{\' e}rsic index $n$. For the former, as introduced in Sect. \ref{sec:CNN}, GaLNet-2 can correct the prediction depending on how round the 2D PSF is from the input local PSF image, according to what was learned in the training phase. For the latter, as discussed above, GaLNet-2 can account for the effect of the smoothing of the profile, especially in the central pixels. 
We stress here that, for both these parameters, GaLNet-1 is also trained to account for the PSF, as this is incorporated in the convolution of the model with a random image from the PSF sample (see Sect. \ref{sec:training_data}). However, when making predictions, GaLNet-1 can only guess on the basis of a ``self-learned'' PSF\footnote{The information of the PSF is encrypted in the mapping between the ground truth labels and the training images which are by definition convolved with the local PSF. Hence, even without knowing the local PSF, GaLNet-1 learns the ``average'' effect of the PSF over the training sample.}, as it does not have information 
on the local PSF (PSF-mismatch, hereafter). This produces a larger scatter on the output parameters because of the well known degeneracies among the S{\' e}rsic parameters (e.g., $R{\rm e}-n$ and the $mag-n$). On the other hand, GaLNet-2, by avoiding the PSF-mismatch using the local PSF, is able to contain the scatter and overall parameter uncertainties (see Sect. \ref{sec:stat_err}).  


We can finally visualize the differences between the two CNNs 
by estimating the residual images obtained by subtracting the 2D S{\' e}rsic model from the original galaxy image. This is done as follows. First, we derive a 2D galaxy model by the estimated parameters issued by the CNN, then convolve this model with the local PSF (i.e. the one we use as input for the GaLNet-2). This PSF convolved model is subtracted by the galaxy image input in the GaLNets: the better the models, the more the residual images look like a (zero) background image.

In Fig. \ref{fig:residual_simulation}, we show some examples of the original simulated cutouts and their residual images from the two CNNs. Most of the time, GaLNet-1 and GaLNet-2 provide acceptable residuals (the first and second rows), even in the presence of companions, partially overlapping systems (second rows). However, in some cases, some bad residuals are found from only GaLNet-1 (third row) or both GaLNet-1 and GaLNet-2 (fourth row). For these cases, most of the bad fits are concentrated in the central pixels 
while the residuals become better outward. We note, though, that the failing cases are generally the ones with bright companion objects that can either affect the closer side of the main galaxy or cannot even be easily deblended. These cases are quite difficult to resolve also with standard methods (see also below).

Finally, we can clearly see that even in the poorer fits, those residuals from GaLNet-2 are generally better than that from GaLNet-1, which is visual proof of the crucial impact of the local PSF on the CNN performances, especially to find the best fit of the galaxy cores.

\subsection{Assessing the statistical errors}
\label{sec:stat_err}
The simulated galaxies, with ``ground-truth" values of the parameters, allow to assess the statistical errors ($e_{\rm s}$, hereafter). In this paper, these will be quantified as the standard deviation of the  scatter between the true values and the predictions
\begin{equation}
    e_{\rm s}^2=\frac{1}{N-1}\sum_i^N(p_i-t_i)^2
\end{equation}
where the $p_i$ and the $t_i$ are the predicted and true values in a given interval of $i=1...N$ galaxies.  

Being computed on simulated galaxies, these errors are referred to idealized objects while, as repeatedly specified before, 
in reality, we will not deal with single S{\'e}rsic profiles. However, these errors are the closest estimate that we can have of the contribution of the 1-S{\'e}rsic ``fitting" procedure to the overall error budget. From this point of view, the statistical errors we derive in this section represent a lower limit of real errors.


For each parameter, in Fig. \ref{fig:test_simulation} we plot the $e_{\rm s}$ as errorbars labelled with the relative error with respect to the mean of the true values in 5 bins. These latter are defined as $\Delta e=e_{\rm s}/{\overline{p}}$, where $\overline{p}$ is the mean of the predicted values in a given bin -- except for $\Delta mag $ that being a logarithmic quantity is defined as \rui{the difference between the predicted and the true magnitude, $e_{\rm s}$}.
Generally speaking, the $\Delta mag$ and the $\Delta e$ of $x_{\rm cen}$, $y_{\rm cen}$, $R_{\rm eff}$ and $q$ for both GaLNets are smaller than $\sim$0.1, meaning that the two GaLNets are rather accurate to estimate these parameters. However, for $PA$ and $n$, most of the $\Delta e$ are between $\sim 0.1-0.2$, slightly worse than that of other parameters. We also clearly see that in each panel the $\Delta e$ of GaLNet-2 are smaller than the corresponding value of GaLNet-1. The differences between the two GaLNets are even larger for $PA$, $q$ and $n$, with an average value of $\sim0.03-0.05$, while for others the differences are generally $<0.2$.

Looking into the variation of the relative errors as a function of the different parameters, we can see that the $\Delta e$ of magnitudes from both GaLNet-1 and GaLNet-2 are generally smaller than 0.06 and stay almost constant in all the range of magnitudes that we have adopted, i.e. from $mag=17$ to $22$. This means that, above a given SNR ($\geq50$), both GaLNets achieve very high accuracy in estimating the total magnitudes, regardless of how luminous they are. The $\Delta e$ of $R_{\rm eff}$ 
are smaller at $R_{\rm eff}<2''$ ($\sim 0.104$ for GaLNet-1 and $\sim 0.085$ for GaLNet-2), and become larger at $R_{\rm eff}>2''$ ($>0.12$ for GaLNet-1 and $>0.1$ for GaLNet-2). 
This is because most of the galaxies with larger $R_{\rm eff}$ also have large $n$. These galaxies are more difficult to model either for standard methods (see discussion in \citealt{LaBarbera2010+SPIDER-I}) and also for GaLNets, as the innermost gradient of the light profiles become very steep and difficult to constrain.
Some further contributions to the scatter can come from 1) the lack of galaxies with large $R_{\rm eff}$ in the training sample, produced by the log--normal distribution (see Tab. \ref{tab:parameters}) and 2) the cutout size (i.e. $12''$ by side), which, in principle, can sample only the central regions of galaxies with large effective radii. 
Since the effect of the cutout size has been shown to not improve the GaLNet's results (see \S\ref{sec:training_data}), the effect of the correlated variation of the  S{\'e}rsic index $n$ and $R_{\rm eff}$ seems to be the most reasonable explanation of the observed errors.


Going to $PA$, the relative scatter $\Delta e$ at $PA<20$ deg and $PA>160$ deg are larger than the other 3 bins in each panel, because, by symmetry, galaxies with $PA\sim180$ deg are equivalent to the ones with $PA\sim0$ deg. This means that the $\Delta e$ of the closer bins to these values does not represent the real PA statistical errors. 
In the other three bins, though,
the $\Delta e$ of $PA$ for GaLNet-1 varies from 0.12 to 0.22, and for GaLNet-2 from 0.084 to 0.20. Considering that a large portion of 
galaxies has $q>0.8$, i.e. they are close to round and it is really hard for them to measure the exact $PA$ value, 10\%-20\% scatter is a satisfactory uncertainty. For $q$, we find the $\Delta e$ for both GaLNets becomes smaller for larger $q$ values. However, this does not indicate that the GaLNet performs worse at large $q$'s. This is because, being the absolute scatter almost constant for all $q$'s, 
the $\Delta e$ becomes larger at small $q$, by definition. It is worth noticing the beneficial effect of the PSFs in GaLNet-2, which can improve its performances  significantly with respect to GaLNet-1 at small $q$. Finally, for S{\'e}rsic index $n$, the scatters are generally larger than that of other parameters, $\sim 0.146-0.184$ for GaLNet-1 and $\sim 0.109-0.131$ for GaLNet-2. In this case, introducing the PSF produces the largest improvement compared to the other parameters.


\subsection{Testing on real data}
\label{sec:testing_on_real} 

\begin{figure*}[htbp]
    \centering
    \subfigure {\includegraphics[width=18cm]{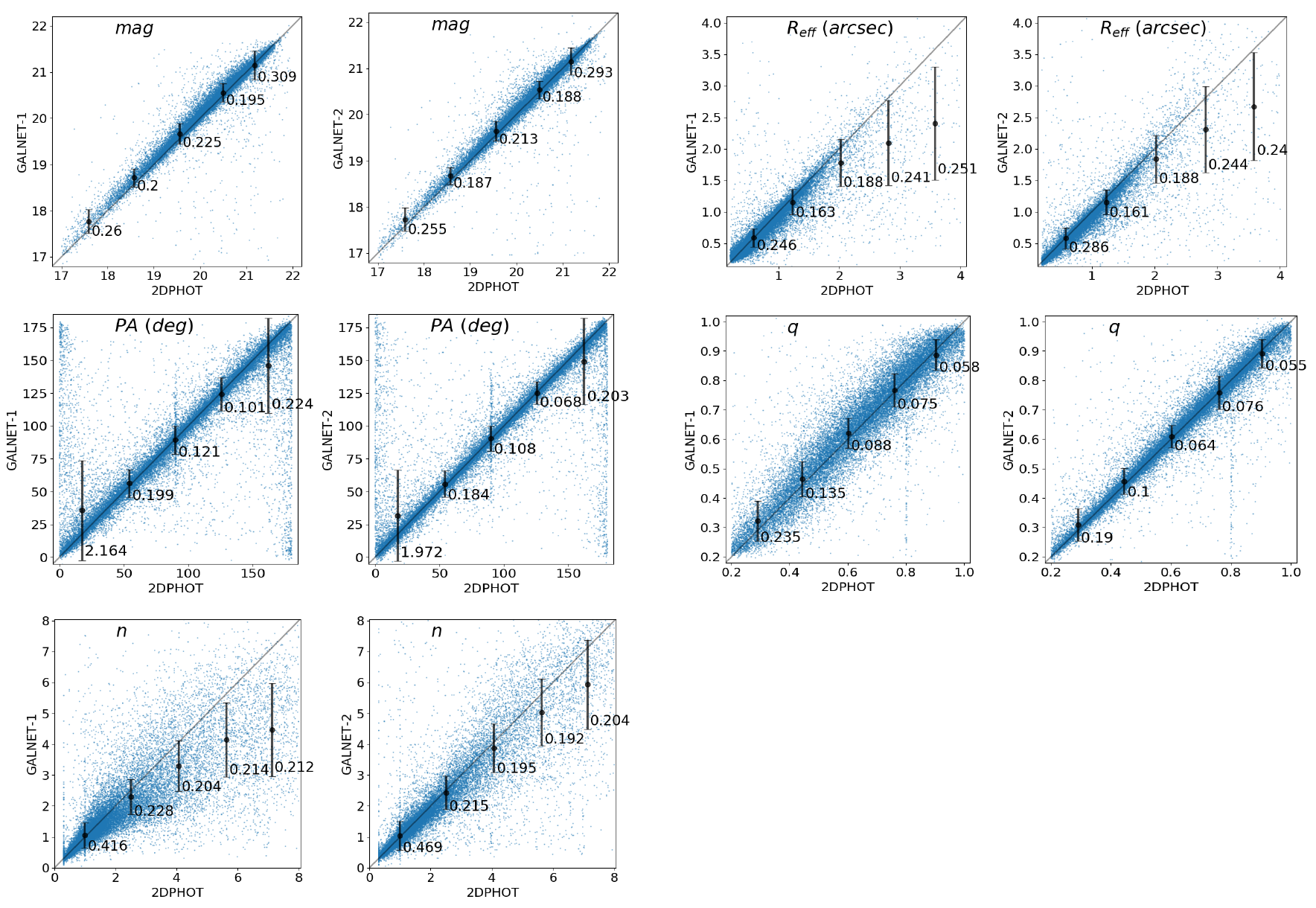}}
    \caption{Comparison between the output of GaLNets and 2DPHOT on real galaxies from KiDS. In each panel, the horizontal axis are the values obtained by 2DPHOT and the vertical axis are the predictions from the GaLNets. {Error bars are the absolute mean errors in each bin, while labels report the absolute errors for $mag$ and relatives errors for others.}}
    \label{fig:real_comparison}
    \vspace{0.1cm}
\end{figure*}

The test on mock galaxies is important to check the presence of systematics and for assessing the statistical errors. Indeed, since the ground truth is error-less, the relative errors discussed in Sect. \ref{sec:stat_err} just contain the internal errors of the measurement method. 

As we finally want to apply the trained GaLNets to real galaxies, we need to check the presence of systematics in more realistic situations, including the impact of unaccounted substructures on real systems. Here we foresee two main categories of ``reality mismatch''. First, a ``model mismatch'': as discussed earlier, for this first GaLNets, we adopt the simplistic choice of a 1-S{\' e}rsic profile. This is possibly accurate enough for a majority of galaxies, but not for all of them. Hence, we need to 
bear in mind that, 
despite how good a method is to fit a simplistic model, 
the best-fit cannot always be optimal, as the real galaxy can have a more complex structure (this is true for both ML and standard tools, see e.g. \citealt{Peng2002+Galfit, Peng2011}).
Second, a ``substructure mismatch'': this comes from the fact that even real galaxies with a dominant 1-S{\' e}rsic light distribution can have uneven components (spiral arms, rings, bars, tilted isophotes, etc.). Of course, there might be systems where both mismatch-types are present. These systems would be rather difficult to fit with a parametric model by general purpose tools, so they are not a matter of concern in this context.
As mentioned in Sect. \ref{sec:intro}, in this first step of the GaLNet development, we will mainly focus on the substructure mismatch and ignore the model mismatch, which will be considered in the next steps of the project.
In particular, we will consider either multi-component parametric models or more general non-parametric models (e.g., \citealt{ Lauer1985ApJS...57..473L, 
Tody1986SPIE..627..733T,
Ciambur2015ApJ...810..120C, Bradley2020zndo...4044744B, Stone2021MNRAS.508.1870S}).


For the test on real galaxies, we use 25\,000 high SNR galaxies randomly selected from KiDS, for which the S{\' e}rsic parameters with 2DPHOT package (\citealt{LaBarbera_08_2DPHOT}) are available (see also \citealt{Roy+18}). 2DPHOT is an automated software environment used for source detection and deep wide-field images analysis. In particular, 2DPHOT can derive structural parameters of galaxies by fitting the images with two-dimensional PSF-convolved S{\'e}rsic models. 
For each galaxy, 2DPHOT can automatically select 2 or 3 nearby sure stars and produce an average 2-dimensional PSF by modeling them with two Moffat profiles (see detail in \citealt{LaBarbera_08_2DPHOT}). This PSF will be convolved by the modelled S{\'e}rsic profile to produce a 2D distribution of the galaxy light. The best fitting parameters of the S{\' e}rsic profile then are obtained by minimizing a given $\chi^2$ function. In the fitting process, the mask images, created by 2DPHOT itself, during the source detection, are also considered.

The real test galaxies are collected according to the following requirements: i) SNR$>50$; 
ii) $17<mag<22$ (in $r-$band); iii) $0.2''<R_{\rm eff}<4''$; iv) $0.1<n<8$; and v) $q>0.2$, which are generally within the parameter intervals of the training data. 
In order to avoid systematics due to the selection and modeling of the ``local'' PSF, as input for the GaLNet-2 we will adopt the PSF models produced by 2DPHOT for the test galaxy sample (see Roy et al. 2018 for details). This has the rationale to strictly evaluate the relative performances of the two tools, but it has the disadvantage that GaLNet-2 will reproduce the same systematics of 2DPHOT in case of poor PSF model occurrences. However we expect to produce independent PSF models for future extensive analyses on real data.

\begin{table}[htbp]
\begin{center}
\caption{\label{tb:R2_real} Statistical properties of real data.}
\begin{tabular}{c c c c c c c}
\hline \hline
CNN model  & $mag$ & $R_{\rm eff}$ & $q$ & $PA$ &$n$ \\
\hline
\multicolumn{6}{c}{$R^2$}\\
\hline
GaLNet--1   & 0.9037  & 0.6666  & 0.6528  & 0.8481  & 0.4274  \\
GaLNet--2  & 0.9166   &0.7035   & 0.7411 & 0.9160  & 0.7488  \\
\hline
\multicolumn{6}{c}{Fraction of Outliers}\\
\hline
GaLNet--1  &0.001& 0.073& 0.239& 0.0152& 0.329 \\
GaLNet--2 & 0.001& 0.065& 0.180& 0.0113& 0.191 \\
\hline
\multicolumn{6}{c}{NMAD}\\
\hline
GaLNet--1   & 0.0040& 0.0497& 0.0668& 0.0327& 0.1242 \\
GaLNet--2  & 0.0034& 0.0259& 0.0437& 0.0140& 0.0786\\
\hline \hline
\end{tabular}
\end{center}
\textsc{Note.} ---Statistical properties of the prediction on real data. From up to bottom we show the $R^2$, the fraction of ourliers and the NMAD for the magnitudes $mag$, effective radius $R_{eff}$, axis ratio $q$, position angle $PA$ and the S{\'e}rsic index $n$.
\end{table}

Differently from the simulated sample, here we cannot know the true parameter values, but we take the estimates from 2DPHOT as the ``ground truth'', assuming that these are unbiased estimates of the intrinsic galaxy parameters. 
Note that this is an idealization. In fact, although 2DPHOT was widely tested with ground-based and HST-data, as well as  simulated galaxies (\citealt{LaBarbera_08_2DPHOT}, \citealt{Roy+18}), some galaxies, e.g. those not well represented by a single S{\'e}rsic profile, and/or in case of poor PSF modeling, might be still affected by unaccounted systematics.

\begin{figure*}[htbp]
\centerline{\includegraphics[width=18cm]{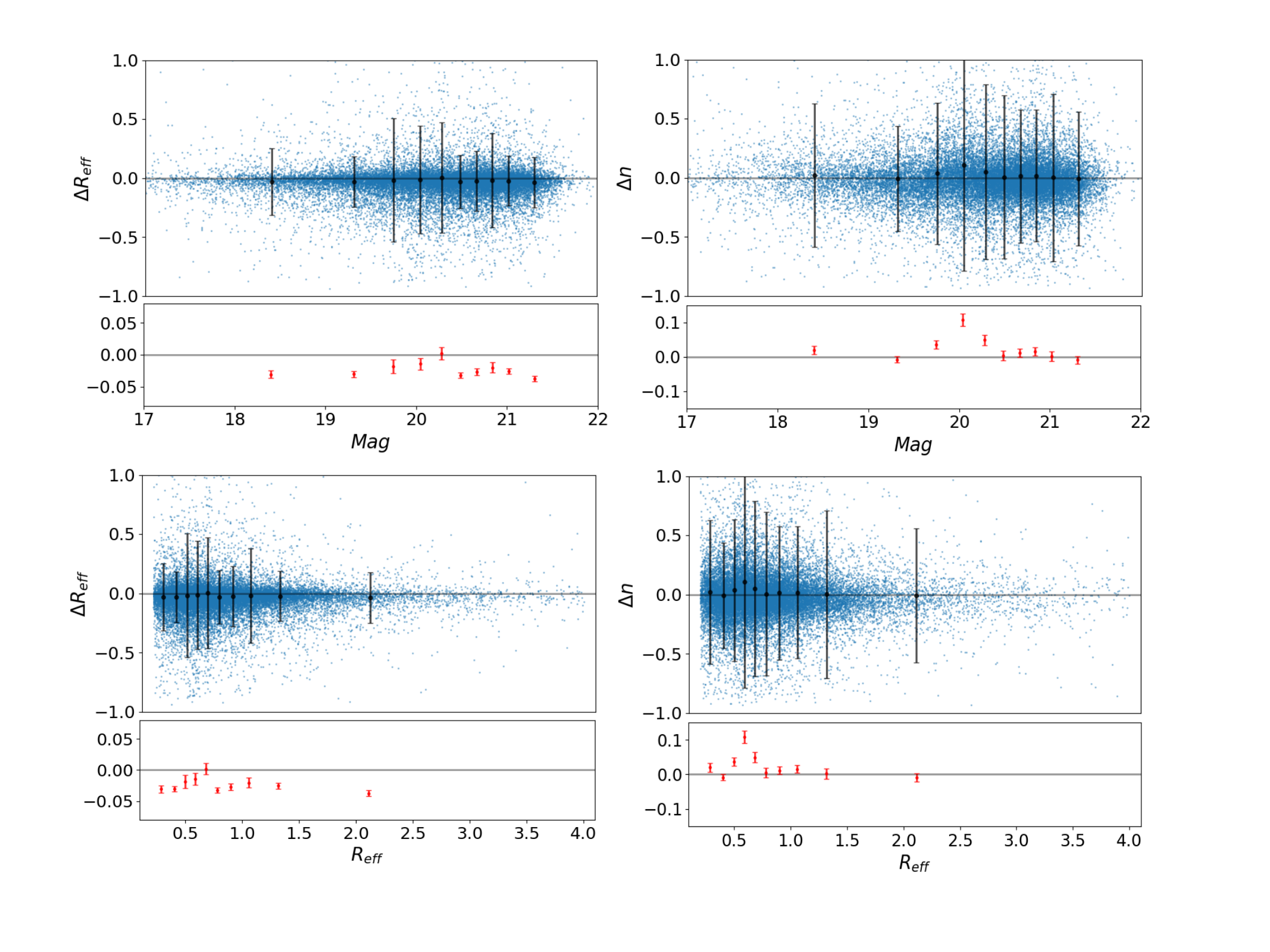}}
\caption{The relative errors of effective radius ($\Delta R_{\rm eff}$) and S{\'e}rsic index ($\Delta n$) against the magnitude ($mag$) and the effective radius ($R_{\rm eff}$). Predictions are from GaLNet-2. In each panel, the blue points are the relative errors and the dark bars show the standard deviation of the errors, while the red points show the mean of the relative errors and the red bars are the standard error on the mean in each bin.}
\label{fig:errors_mag_reff}
\vspace{0.5cm}
\end{figure*}

However, being based on completely different approaches it is unlikely that both methods suffer the same systematics. 
Hence, from their comparison we can learn more about their relative accuracy,
particular whether the GaLNets provide parameter estimates consistent with standard fitting approaches.

The comparison between the parameters $mag,~R_{\rm eff},~ q,~ PA$, and $n$, obtained by 2DPHOT and the two GaLNets, are shown in Fig. \ref{fig:real_comparison}. 
For these one-to-one relations, the $R^2$, outliers and NMAD are reported in Table \ref{tb:R2_real}. For all 5 parameters, we generally see a fair agreement between the estimates of the two tools, being all data concentrated around the one-to-one line and the $R^2$ parameter larger than $0.7$. 
Looking at the comparison of the individual 
GaLNets we see that, as for the test sample, also for real galaxies 

GaLNet-2 shows a tighter consistency with the ``ground truth'' estimates than GaLNet-1,
especially for $q$ and $n$. 
However, magnitudes and effective radii of galaxies are still reasonably well constrained from GaLNet-1, showing that the PSF is important for these parameters mainly to reduce the uncertainties, but that the mean estimates over large numbers  are fairly unbiased.

\begin{figure*}[htbp]
\centerline{\includegraphics[width=18cm]{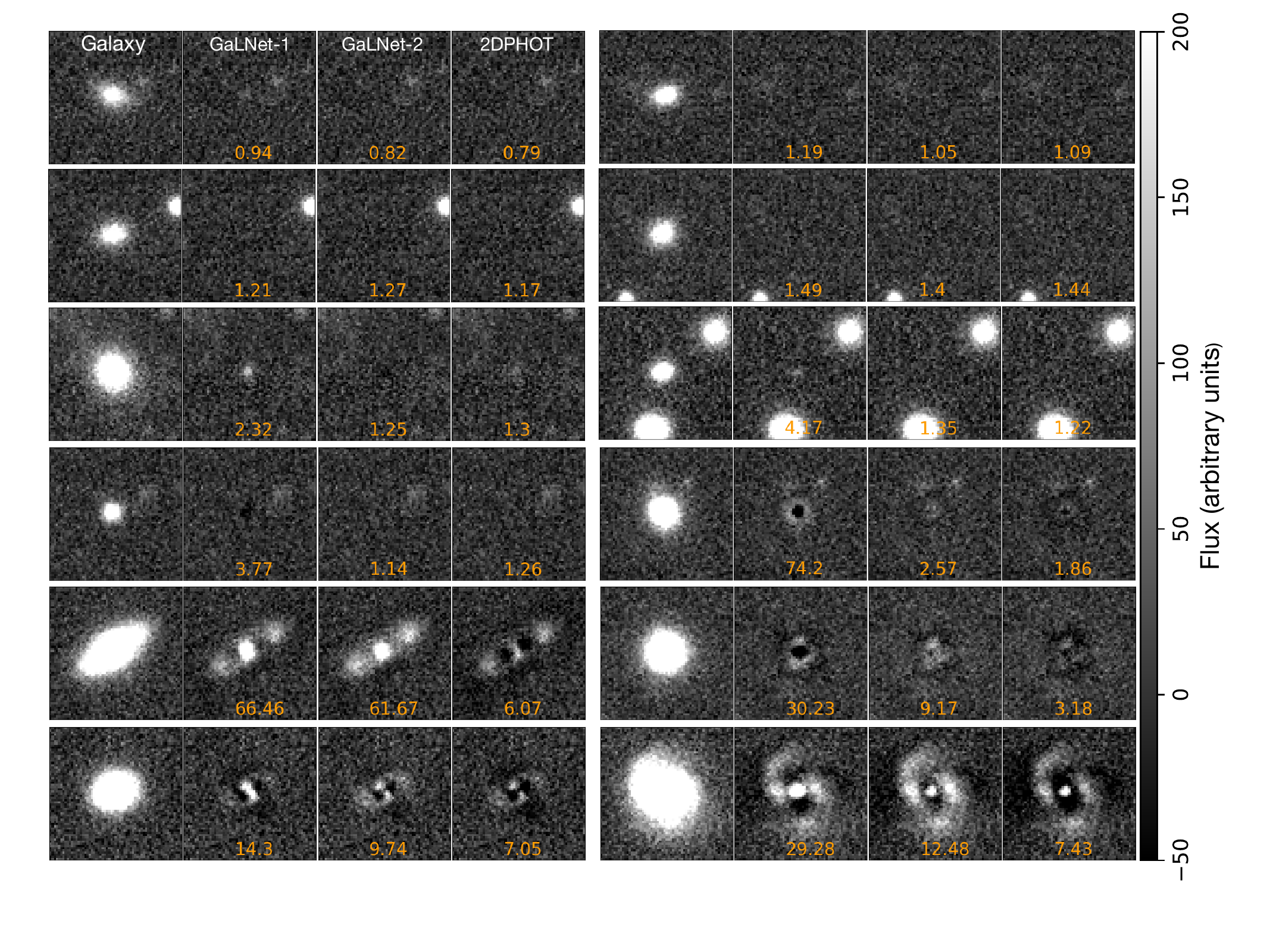}}
\caption{Residuals of real galaxies used for testing the networks. In each panel, the first is the $r-$band image of real galaxy, the second and third are the residual obtained from the GaLNet-1 and GaLNet-2 predictions, respectively, and the fourth is the residual obtained from 2DPHOT fitting. At the bottom of each residual image, we report the a posteriori reduced $\tilde{\chi}^2$.
}
\label{fig:real_residuals}
\vspace{0.5cm}
\end{figure*}

\begin{figure*}[htbp]
\centerline{\includegraphics[width=18cm]{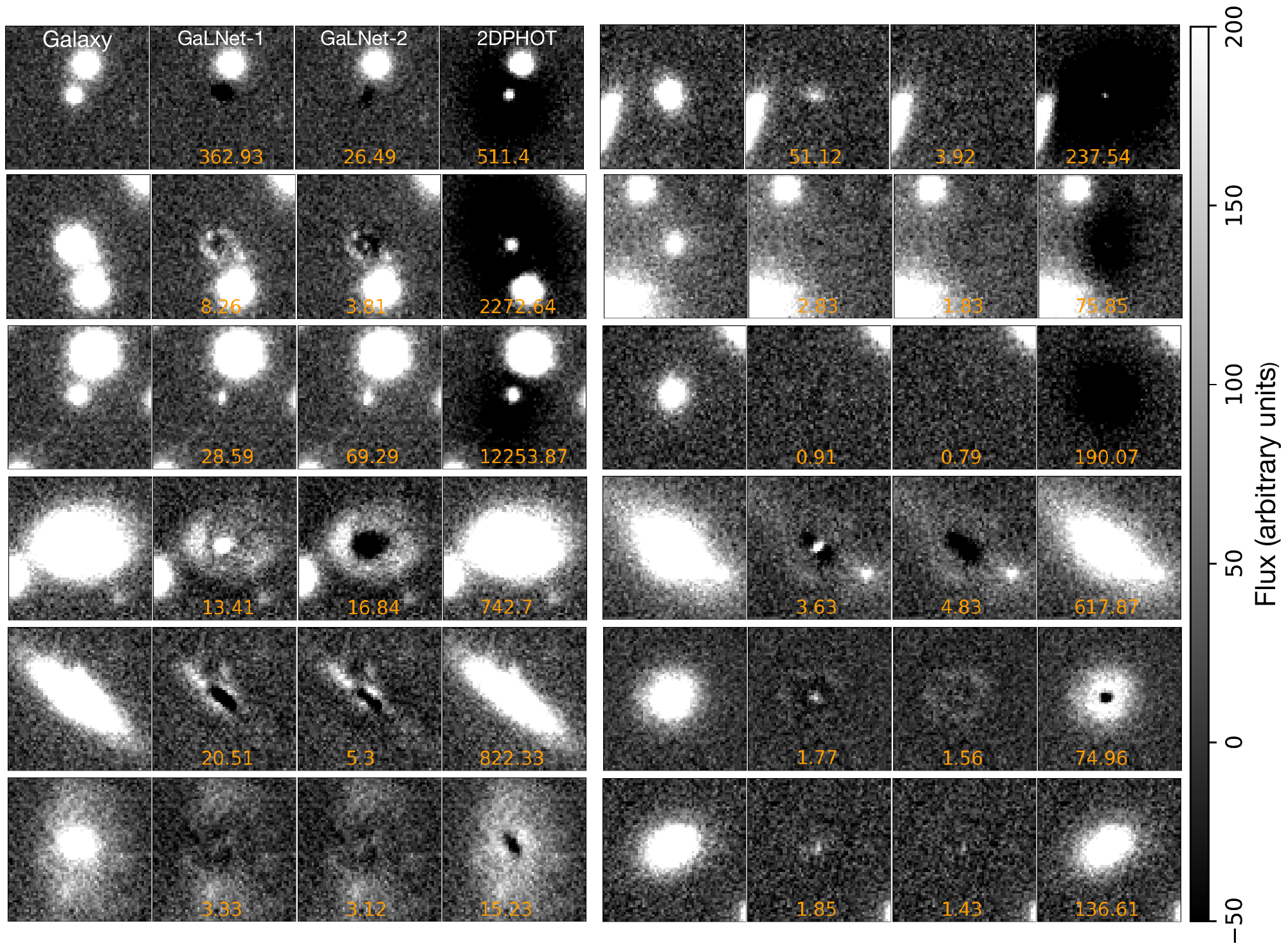}}
\caption{GaLNet vs. 2DPHOT modeling for outliers in the magnitude estimates. These are galaxies expected to have failed to converge to an acceptable model. Panels are arranged as in Fig \ref{fig:real_residuals}.}
\label{fig:residual_magnitude}
\vspace{1cm}
\end{figure*}

Compared with simulated data, the $R^2$ values are poorer overall than those  measured in Sect. \ref{sec:testing_simulation}, showing a degradation of the performances with respect to the test sample. 
The larger scatter seen in Fig. \ref{fig:real_comparison}, is not entirely due to larger uncertainties of the GaLNets, but rather by the combination of the statistical errors from the two different measurement procedures. Indeed, differently from the test sample, where the ground truth values were error-less, here the 2DPHOT estimates carry their own intrinsic error. 
\nic{If the two methods had similar Gaussian statistics, and the uncertainties were uncorrelated, we might expect the scatter}
in Fig. \ref{fig:real_comparison} to be $\sim\sqrt{2}$ times larger than the one in Fig. \ref{fig:test_simulation}. For instance, the mean scatters reported in Fig. \ref{fig:real_comparison} for GaLNet-2 is a factor $\sim2$ larger than the one in Fig. \ref{fig:test_simulation} for the $R_{\rm eff}$, and a factor $\sim1.6$ larger for $n$-index.
This confirms that the GaLNets do not perform significantly worse on real galaxies, considering that some extra uncertainties on both the GaLNets and 2DPHOT come from the substructure mismatch (i.e. the fact that real galaxies possess extra substructures) or the model mismatch (i.e. the fact that real galaxies might have multiple components, like a bulge and a disk), as discussed in Sect. \ref{sec:testing_on_real}. Moreover, as noted above, both GaLNets and 2DPHOT were used to measure the same galaxies, implying that their uncertainties are not necessarily uncorrelated.
Before we move on, we need to point out the presence of some vertical features in  Fig. \ref{fig:real_comparison}, suggesting the presence of some systematics in the 2DPHOT estimates. These are found for $PA\sim90$ deg and $q\sim0.8$ for both comparisons with GaLNet-1 and GaLNet-2. 
We have visually checked a sub-sample of these outliers and found that galaxies at the $PA\sim90$ deg always have a very round shape. 
Since for these systems, it is hard to define the $PA$, 2DPHOT tends to give the $90$ deg values, close to the initialization value. As for the second feature at $q\sim0.8$, we notice that the galaxies always have close pairs, and the light from the pairs has a strong influence on the central galaxies. In some cases, again, 2DPHOT cannot optimize the $q$, and finally use the initial guess as the final output.

In Fig. \ref{fig:errors_mag_reff}, we show the relative errors (top panels) and the standard errors of the mean (bottom panels) of the effective radius and the S{\'e}rsic index for the GaLNet-2 against the magnitude ($mag$) and the effective radius ($R_{\rm eff}$), as representative of the behaviour seen in Fig. \ref{fig:real_comparison}. These are defined as $\Delta R_{\rm eff}=(R_{\rm eff}^{\rm GN2} - R_{\rm eff}^{\rm 2DPH})/R_{\rm eff}^{\rm 2DPH}$, and $\Delta n=(n^{\rm GN2} - n^{\rm 2DPH})/n^{\rm 2DPH}$, where we have used the apex GN2 and 2DPH, for GaLNet-2 and 2DPHOT, respectively.
From the two plots in the upper row, we 
see no clear dependence of the 
$\Delta R_{\rm eff}$ and $\Delta n$ with the total magnitude, although the bulk of the scatter seems to increase while the galaxies become fainter.  
In the left panel of the upper and bottom row,
we see that GaLNet-2 predicts effective radii that are systematically smaller than 2DPHOT (as also seen in Fig. \ref{fig:real_comparison}). Statistically, these seem to be marginally significant, as about half of the standard deviation of the mean
(red errorbars in the bottom panels) are generally deviating by more than $2 \sigma$ from the zero value in the different $mag$ and $R_{\rm eff}$ bins.
To investigate in details the reason for this discrepancy is beyond the purpose of this paper, as this would involve the use of a third tool to cross-check the systematics. This will be explored in more detail in future analyses.
The scatter of the $\Delta R_{\rm eff}$ seems to be also invariant
with $R_{\rm eff}$.
Finally, in the right panel of the bottom row, we show $\Delta n$ against $R_{\rm eff}$. Here we see no clear systematics and fairly similar errors (around 40\% and 50\%), except for the very small $R_{\rm eff}$. Eventually, for these small galaxies, where the $R_{\rm eff}$ is sampled by $\sim$ 2 pixels by side, {both 2DPHOT and the GaLNets can still measure the slope of the light profile with no systematics, but with larger uncertainties}. The absence of systematics is particularly encouraging, as this would imply unbiased information on large datasets in this particular region of the parameter space. 

In Fig. \ref{fig:real_residuals} we show some examples of the residuals from the model subtracted images. These are derived as in Sect. \ref{sec:testing_simulation}. 
For each target we show the raw image of the original $r-$band image of the galaxy (first column from the left), and the residuals from GaLNet-1 (second column), GaLNet-2 (third column) and 2DPHOT (fourth column). 
For most of the galaxies, as the ones shown in the first two upper rows, the performance of the 3 tools are fairly good, with almost no residuals. 

To quantify this we have computed an {\it a posteriori} reduced $\tilde{\chi}^2$, defined as
\begin{equation}
\tilde{\chi}^2=\sum_i \frac{(p_i - m_i)^2}{\sigma_{\rm bkg}^2}/{\rm dof} 
\label{eq:chi2}
\end{equation}
where $p_i$ are the observed pixel counts of the galaxy within the effective radius, $m_i$ the model values, the $\sigma_{\rm bkg}^2$ the background noise and dof$=($n. of pixels$-$n. of fit parameters). 
This definition of $\chi^2$ for both GaLNets and 2DPHOT is different from the one used in \cite{Roy+18} for 2DPHOT, and serves here only as an overall estimate of the relative goodness of the final GaLNets' fits and compare these with 2DPHOT. These are reported in Fig. \ref{fig:real_residuals}, where one can see that smaller $\chi^2$ correspond, in general, to smaller local residuals.

In the same Fig. \ref{fig:real_residuals}, we also see cases where the GaLNet-1 cannot perform as better as the other two methods, as we can spot darker/brighter dots in the center due to over/under-subtraction, respectively (see e.g. third and forth rows). To our surprise, even if the CNNs have been trained on smooth one component profiles, in case of bright real substructures like spiral arms, bars, cores, rings, the two GaLNets tend to predict rather reasonable S{\' e}rsic profiles, without catastrophic residuals.
This is not what often happens to standard tools based on $\chi^2$ minimization or Markov Chain MonteCarlo (MCMC), which sometimes diverge and return failing results. As an example, we have checked the tail of solutions located at the lower-right corner of the total magnitude plot (e.g. 2DPHOT: $20<mag<22$, GaLNets: $17<mag<19$, in Fig. \ref{fig:real_comparison}) and the other outliers toward the top-left corner in the same plots. Such large outliers are unexpected for the $mag$ prediction as this is one of the tightest constrained parameters both from the 2DPHOT and the GaLNets.


In Fig. \ref{fig:residual_magnitude} we show the residuals of some of these ``outliers''. The three upper rows are six outliers from the top-left corner of Fig. \ref{fig:real_comparison}, i.e. systems for which 2DPHOT measures a bright magnitude and the two GaLNets predict they are faint ($r>21$) objects. It is clear, in all cases, that 2DPHOT over-predicts the magnitude, ending in a wide dark area in the center of the residual image. This is due to a mix of situations: 1) the presence of close objects; 2) the presence of a steep peak in the galaxy center. 
In this latter case (usually a small bulge, but also an overlapping compact object, see e.g. top-left)
the two GaLNets focus on the overall 2D light gradients and ignore the central peak, hence producing a rather satisfactory fit. In all other cases, the presence of close systems could make the segmentation images produced by S--Extractor (\citealt{Bertin_Arnouts96_SEx}) and used by 2DPHOT to be not very accurate. Therefore, it becomes more difficult for 2DPHOT to find a correct fit, as the contribution of exceeding light in the outskirts pushes it to converge toward high-$n$ models.


Going to the bottom three rows in Fig. \ref{fig:residual_magnitude}, i.e. systems for which 2DPHOT measures a faint magnitude and the two GaLNets predict they are bright ($r<19$) objects, the situation is reversed, but yet it is mainly the standard model failing. Here, the problem seems to be the presence of close companions that 2DPHOT fails to mask, leading to some under-fit.
The two GaLNets, on the other hand, seem to nicely catch the main smooth component of the galaxies, correctly leaving the peaked components behind.

There are two aspects of this test that are particularly significant. First, the CNN approach does not generally produce catastrophic events, at least in terms of total magnitude,
while standard methods sometimes do. 
Also, CNNs always converge while standard methods sometimes do not.
Second, CNNs seem to learn how to correctly ``weight'' the main component of a galaxy, as they learn from the training sample that this can be mixed to other secondary substructures, either close galaxies or stars, or small embedded systems (see e.g. Fig. \ref{fig:simulation}). This natural ``penalization'', which is intrinsic in the learning process, turns out to solve very efficiently most of the issues related to standard tools in terms of over/under-fitting of the data.   
{\it These are not negligible capabilities that represent, in our view, a strong advantage of the ML approaches}, provided that the structural parameters of the targets fall within the properties of the training sample.

\begin{figure*}[htbp]
\centerline{\includegraphics[width=18cm]{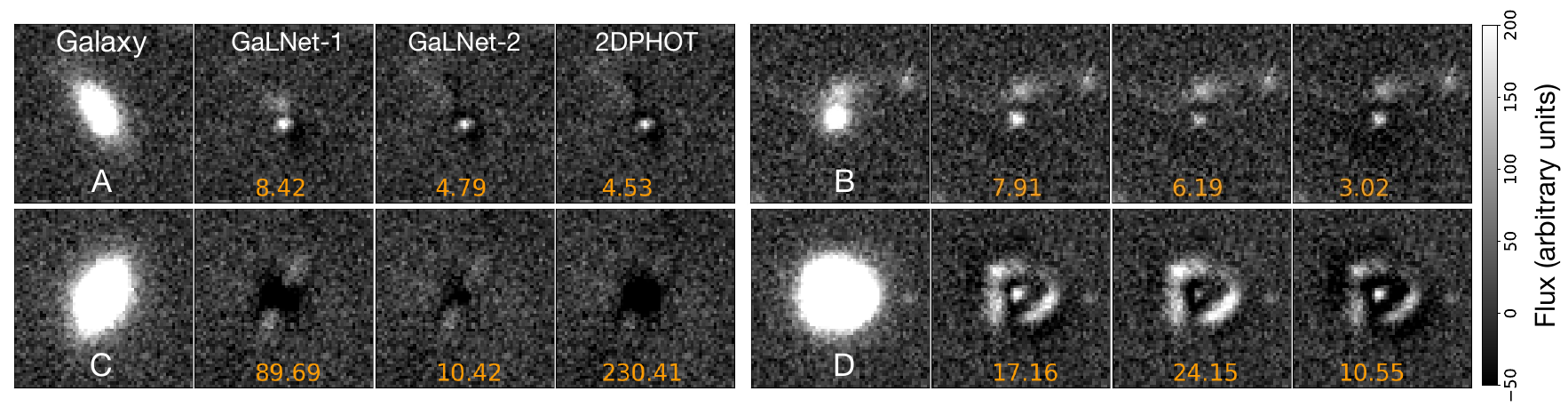}}
\caption{\rui{We show some special features, like compact cores (A), interactions (B), residual disks (C), and  pseudo-lensing (D),} in the residuals of real KiDS galaxies. Panels are arranged as in Fig \ref{fig:real_residuals}.}
\label{fig:special_feature}
\vspace{0.5cm}
\end{figure*}

Of course, in all this analysis above we have left in the background the model-mismatch imposed by our initial choice of the single S{\' e}rsic profile. Most of the catastrophic outliers, a minor fraction amounting to $\sim1\%$ of all galaxies analyzed in the comparison shown above, could be resolved by adopting of more realistic two-component models for both the standard and CNN approaches. This will be addressed in forthcoming papers. Here we just stress that GaLNet-2, in particular, is able to capture the main component, when evident, and leave behind other substructures, 
like clear spiral arms 
(see Fig. 6) 
as well as more peculiar features, like compact cores, interactions, residual disks, pseudo-lensing configurations, etc. (see Fig. \ref{fig:special_feature}).

Noticeably, 
robust 1-S{\' e}rsic analyses are 
extremely valuable {\it per se}, as they provide crucial parameters for a large sample of galaxies (total luminosity, sizes and $n-$index as proxies of the morphology) with enormous science implications.




\section{Discussion and perspectives}
\label{sec:discussion}
In this section, we discuss in some more detail the possible reasons for the different performances of GaLNet with respect to standard 2D fitting approaches and roughly estimate the gain in terms of computational time. 
We will also discuss the perspectives in terms of future developments of the GaLNet tools. 

\subsection{GaLNets vs. standard methods}
\label{sec:chi2}
In Sect \ref{sec:testing} we have compared the results of the two GaLNets against each other, and against 2DPHOT, and discussed the relative scatter as a measure of their internal statistical errors. We have shown that the performances of GaLNet-2 are generally better than the ones of GaLNet-1 both in reproducing the ground truth of the simulated (test) sample, and the estimates from 2DPHOT of the KiDS (real) galaxies.

Here we briefly look into a more quantitative comparison between the different tools and discuss if there are systematic differences and what is their origin. 

\begin{figure*}
\centerline{\includegraphics[width=18cm]{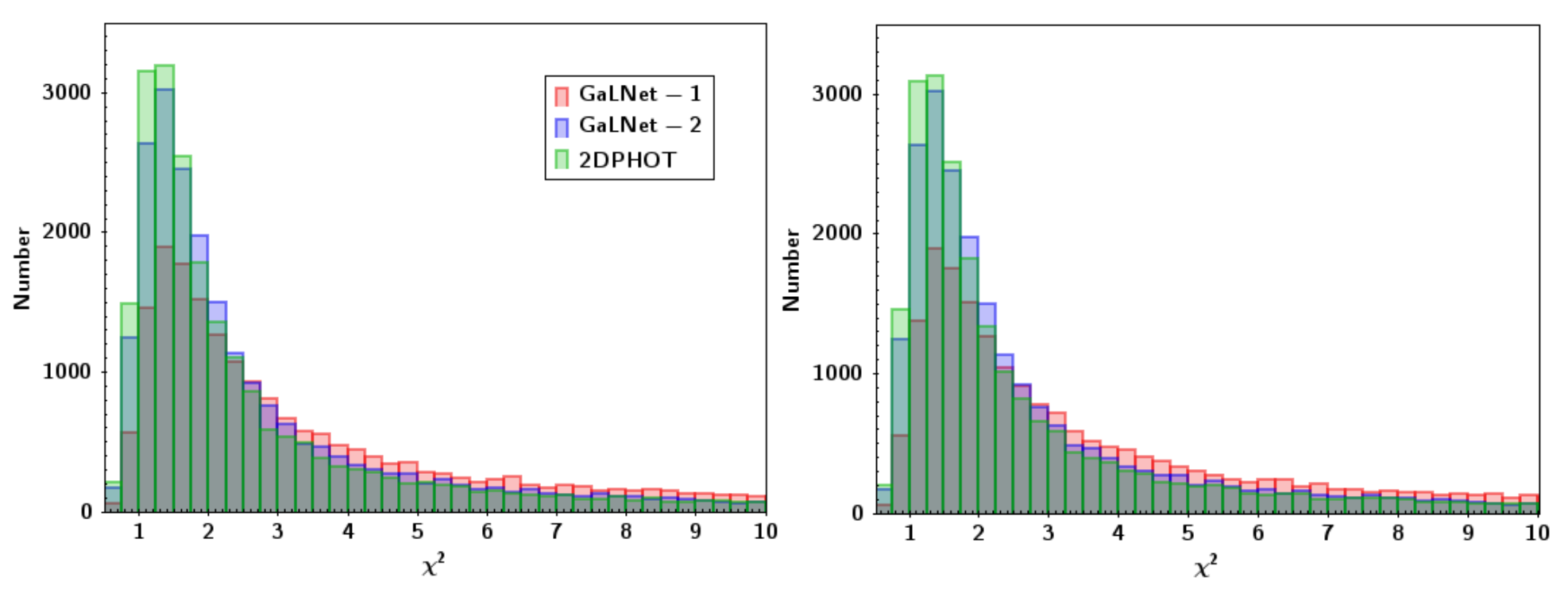}}
\caption{$\chi^2$ distribution of GaLNet-1, GaLNet-2 and 2DPHOT. The ones in the left are obtained using pixels  within the individual method effective radius, while the ones in the right are obtained from pixels inside the same aperture defined by the effective radius from GaLNet-2. The x-range is cut to $\chi^2=10$ for sake of readability and the distribution is almost flat outside the window.
}
\label{fig:chi2}
\vspace{1cm}
\end{figure*}

In Fig. \ref{fig:chi2}, we show the distribution of the $\tilde{\chi}^2$ defined in Eq. \ref{eq:chi2}, for the two GaLNets and 2DPHOT. {To make a uniform comparison among the different tools, we show both the  $\tilde{\chi}^2$ obtained using pixels within the individual method effective radius and the one obtained from pixels inside the same aperture, defined by the effective radius from GaLNet-2, for all of them. As we can see, the two definitions are almost equivalent at all $\tilde{\chi}^2$, as they do not produce appreciable differences in their distributions. The major relative differences are possibly present at 
larger $\tilde{\chi}^2$, where the differences of the estimated parameters are larger, including strong outliers, as discussed in the previous sections. 
} 
Generally speaking, we see that the $\tilde{\chi}^2$ of 2DPHOT tend to be better than the one produced by GaLNet-2 because the distribution is peaked slightly toward a smaller $\tilde{\chi}^2$ (closer to 1). GaLNet-2 has a small excess of $\tilde{\chi}^2$ around 2-3, while the two distributions are identical for $\tilde{\chi}^2>3$. GaLNet-1, on the other hand, is more broadly distributed with a large number of systems at $\tilde{\chi}^2>3$ than the other two algorithms.
As already discussed in Sect. \ref{sec:testing_simulation} and \ref{sec:training_data}, the main reason for this overall underperformance of GaLNet-1 is the unaccounted local PSF. 

\begin{figure*}[htbp]
\centerline{\includegraphics[width=18cm]{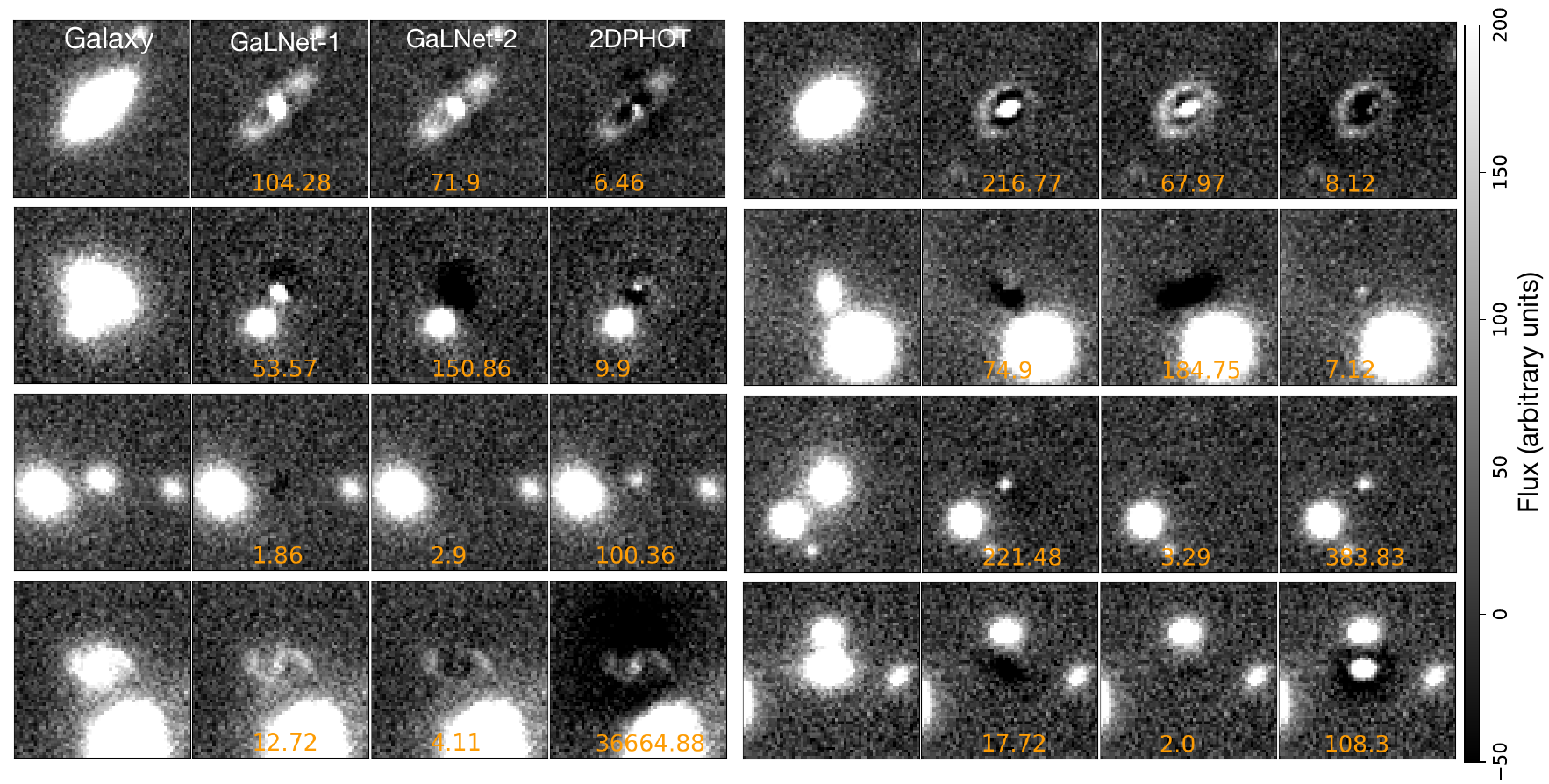}}
\caption{special cases of the model of real KiDS galaxies. We show residuals with $\tilde{\chi}^2_{\rm GN2} >> \tilde{\chi}^2_{\rm 2DPH}$ in the top two rows and residuals with $\tilde{\chi}^2_{\rm GN2} << \tilde{\chi}^2_{\rm 2DPH}$ in the bottom two rows. Panels are arranged as in Fig \ref{fig:real_residuals}.
}
\label{fig:special_case}
\vspace{0.5cm}
\end{figure*}

Going to the comparison between GaLNet-2 and 2DPHOT, we have observed that the former performs sensitively worse than the latter, i.e. $\tilde{\chi}^2_{\rm GN2}>>\tilde{\chi}^2_{\rm 2DPH}$
in two major circumstances. First, when there are bright substructures in the center (e.g. spiral arms, rings etc., see Fig. \ref{fig:special_case} top two rows): in this case, GaLNet-2 tends to predict a flatter profile and underfit the outer regions (hence leaving bright residuals in the galaxy outskirts). Second, when there is a bright companion, GaLNet-2 tends to overfit the model in the center and produce darker regions and a worse $\tilde{\chi}^2$.

Finally, there are cases where 2DPHOT performs much worse than GaLNet-2, i.e. $\tilde{\chi}^2_{\rm 2DPH}>>\tilde{\chi}^2_{\rm GN2}$ (see Fig. \ref{fig:special_case} last two rows). In around 50\% of these cases, we have seen bright sources around the main galaxy, which is generally a compact one. Here 2DPHOT is likely unable to mask out the companion galaxies, and this might produce flatter tails in the measured profile which is fitted with a larger $R_{\rm eff}$ and/or smaller $n$, producing an underfitting of the central regions that remain unsubtracted in the residual images. GaLNet-2, on the other hand, being insensitive to the outer regions, 
can fit better the main galaxy component.
For the other 50\% of the cases, we see a mix of situations with no clear pattern, including rather crowded regions, bright large galaxies and strong blended sources, where both 2DPHOT and GaLNet-2 can either perform better or worse, rather unpredictably.
Finally, we have a mixed situation with $\tilde{\chi}^2_{\rm 2DPH}<\tilde{\chi}^2_{\rm GN2}$, or vice versa, and both $\tilde{\chi}^2>>1$, where, looking at the residuals, it is hard to really appreciate sensitive differences. 

Overall, these detailed comparisons above give useful indications about 
improvements to implement for the next GaLNets. For instance, by increasing the number of cutouts of the background sample with bright galaxies in, 
CNN will eventually learn 
how to predict the parameters in the presence of a more severe galaxy-galaxy blending (Sect \ref{sec:future}). 

However, according to the current GaLNets vs. 2DPHOT results, the broad conclusion from Fig. \ref{fig:chi2} is that GaLNet-2 has a predictive power comparable to traditional methods. 

The computational speed, instead, 
is a major strength of GaLNets.
This can depend on many factors including the performance of the hardware underuse. However, to give a scale of the relative computing times, we have made a statistic
for individual galaxy from the 2DPHOT run, which takes a few seconds for the 2D fitting, and a similar time 
for the preliminary 2D fitting that 2DPHOT performs to determine the initial guess parameters (\citealt{LaBarbera_08_2DPHOT}). As a comparison with another standard 2D fitting tool, we have checked the average performances of Galfit on a bunch of KiDS galaxies using the 2DPHOT parameters as a first guess and measured on average $\sim6$s/galaxy computing time. Even ignoring the time needed for the first guess step, the standard tools need the order of seconds to converge. On the other hand, we have seen that 
the CNNs need only $\sim0.04$s/galaxy running on standard CPUs and about a factor of 10 faster  running on
commercial GPUs on a laptop. I.e., the GaLNets can work at a speed which is a factor of $\sim$250 to 1\,500 with respect to standard tools, with hardware which is about 4 times cheaper. If we include the necessary of optimisation step of the initial guess parameters, which for the CNN is incorporated in the training step and is negligible in time for a billion galaxy sample, then the GaLNets are overall a further factor of 10 faster than the standard tools.

With these exceptional computational capabilities, a billion galaxies in a ten of optical + near-infrared bands, like the sample we will collect by combining Rubin/LSST, EUCLID and CSST, will be doable for GaLNets in a week's time on a mid-sized GPU server. This makes GaLNets, and other similar ML tools (e.g. DeepLegato) a very promising solution for the next generation large sky surveys. 

\subsection{Future developments}
\label{sec:future}

In this section, we first discuss the possibilities for improving the performances of the current GaLNetS. Then we will discuss how to expand their applicability to solve some issues related to the structural parameter measurement, for instance, the deblending (as anticipated in Sect \ref{sec:chi2}). 

The first line of the GaLNet improvements is related to the treatment of the PSFs in the training sample simulations. In this process, randomly selected PSFs were used to convolve the S{\' e}rsic profiles. Hence, the PSFs of the central galaxies are different from the local ones of the surrounding galaxies. This can be a problem for the GaLNet-2 to learn how to deblend two or more galaxies as this will not reflect the real situation where there is a unique PSF. This PSF mismatch will be solved in the future, by first selecting a ``PSF sample" and then assembling  the ``background sample" in the vicinity of the selected PSF, hence building ``PSF-background pair'' sample. In this case, when a galaxy is simulated in the center of the background image, the properly paired PSF is convolved with its S{\' e}rsic profile. The PSF can be either provided as a Moffat model, as discussed in Sect. \ref{sec:training_data}, or can be possibly provided as the image of the representative star used to estimate the local PSF.  

As a second improvement, we will extend the application to fainter galaxies with lower SNR. This paper has only focused on high ``average'' SNR galaxies. 
{However, the lower SNR galaxies account for most of the observed targets in galaxy surveys, and they usually have lower masses or are located at higher redshift. These two properties are valuable for the studies of galaxy structure, formation and evolution and for many scaling relations. 
Being more sensitive to the central regions of the galaxy, as discussed in Sect \ref{sec:training_data}, \nic{we will check whether GaLNets may perform reasonably well with lower average SNRs.}}

As a further improvement, thanks to their computational speed, the GaLNets will make it possible to derive the light profiles for multi-band images. In particular, 
we can train a CNN to simultaneously predict all the parameters from all bands. This can help to prevent single-band outliers as one can expect a smooth continuity among the structural parameters in different bands (see e.g. the magnitude outliers in Fig. \ref{fig:residual_magnitude}).
With such a regular multi-band fitting, one can expect to reconstruct the 2D model of the galaxy spectral energy distribution (SED) for stellar population analysis.

Finally, we will implement the multi-S{\' e}rsic profiles in the training sample. Here we expect the degeneracies among the parameters to be increased by the presence of multiple components, but we will possibly take advantage of the multi-band simultaneous fitting above to break some of these degeneracies, as the different components might look different in the different bands.

One missing feature of the current GaLNets is that, as all CNNs, they cannot provide the errors on parameters of individual galaxies (but statistical errors can reasonably be derived from the test samples).
Possible solutions here can be the Bayesian neural networks (e.g. for strong lensing model, \citealt{Levasseur2017} ) or the adoption of some hybrid algorithms, e.g. using the prediction from CNNs as an initial guess for a standard Markov Chain Monte Carlo (MCMC) code (e.g. \citealt{Pearson+21}). 

To conclude this section, we want to list a few further applications that can be performed by the GaLNets, with some simple variations on the training methods. For instance:
\begin{itemize}
    \item[1)] Star/galaxy separation. Using the information of the shape and the internal light profiles, we expect that training the GaLNets on simulated PSF and galaxies will enable the CNNs to work as a classifier to assign a probability to be a star to each modelled object. This is a rather standard task that 
    has been implemented with even sophisticated standard or ML techniques 
    (e.g. \citealt{LaBarbera_08_2DPHOT, Khramtsov2019A&A, Muyskens2021arXiv}).
    However, integrating this step in the analysis will make the GaLNets capable of autonomously selecting the stars for the PSF modeling, hence making a step forward in the full automatization of the tool.
    \item[2)] Deblending. As discussed in Sect. \ref{sec:chi2}, we can improve the capability of the CNN to make models of galaxies blended in close companions. This might have some limit (in terms of accuracy) for a general purpose tool, where one expects the majority of the systems to be almost isolated galaxies. On the other hand, we can specialize a GaLNet only on deblending systems, focusing on a smaller number of relevant parameters (e.g. total luminosity and half-light ratio), but trying to optimize the deblending performance.
\end{itemize} 

\section{conclusions}

With the upcoming facilities for wide-field sky surveys, we will enter a new era
in which an enormous amount of data will be available to study the life of the Universe and the evolution galaxies therein in very detail. The physical processes driving this evolution are heavily connected to galaxy properties, like mass, size, colors and color gradients, shape, internal kinematics etc., and are dependent on the epoch the galaxies are observed, i.e. their redshifts. 

Measuring the structural properties of millions to billions of galaxies is a privilege to do great science, but it will represent a major challenge for the different scientific communities. Not to mention other complex and time-consuming tasks like source deblending and star/galaxy separation.
In preparation for the upcoming dataflows, there is a significant effort from the community to investigate and test new techniques, based on machine learning, to perform these tasks efficiently.
In some specific applications, ML tools are in a rather advanced stage: e.g. photometric redshifts (\citealt{Sadeh2016}, \citealt{Kind2013}) and strong gravitational lens classifiers (\citealt{Petrillo+17_CNN, Petrillo+19_LinKS}, \citealt{Li+20_KiDS, Li+2021+DR5lens}).

There are, though, other applications that are still at the pioneering stage. This is the case of the galaxy structural parameter measurement, which is a major outcome expected to be provided by these billion galaxy surveys. To start filling this gap, we have built two Convolutional Neural Network (CNN) tools to fitting the surface brightness profiles of galaxies from seeing-limited ground-based observations. One network is fed with only galaxy images (GaLNet-1) while the other uses galaxy images together with the local PSF measured in the galaxy surroundings (GaLNet-2). The two CNNs are trained with 200\,000 mock galaxies created by adding PSF-convolved simulated S{\'e}rsic profiles on top of randomly selected cutouts from public KiDS images. The outputs of the CNNs include 7 parameters used to describe the S{\'e}rsic profiles of the galaxies: the coordinates of the center position $x_{\rm cen}, y_{\rm cen}$, magnitude $mag$, effective radius $R_{\rm eff}$, axis ratio $q$, position angle $pa$ and S{\'e}rsic index $n$. We have tested the CNNs on both mock data and real galaxy cutouts collected from KiDS. 

We have found that the two CNNs can
predict values for the structural parameters, which match well with the input values of the simulated galaxies. We have also shown that, when applied to real KiDS galaxies, the predictions are quite consistent with the best-fit values determined from standard PSF-convolved 2D S\'ersic profile fitting procedures (i.e. 2DPHOT, \citealt{Roy+18}).
Specifically, the CNN using the local PSF (GaLNet-2) performs very similarly to 2DPHOT, while the CNN trained with only the galaxy images (GaLNet-1) performs slightly worse, albeit still acceptable. These results demonstrate that the GaLNets are a promising class of regression CNN to perform an accurate analysis of surface brightness of galaxies. We have also shown that they are fast, up to 1000 times faster than standard tools based on standard statistics like $\chi^2$ minimization, like 2DPHOT, which is comparable, if not faster than other standard tools based on likelihood maximisation or MCMC.

Next implementations of GaLNets will add complexity into the ground truth, e.g. use multiple-component models, substructures and obtain simultaneous predictions in multiple bands. Moreover, we can further generalize them to perform other kinds of difficult tasks, as source deblending or star/galaxy separation. 

For all these reasons, GaLNet will represent a unique and unprecedented tool to perform in a very efficient way a series of tasks on datasets from future ground- and space-based facilities (e.g. HSC DES, Rubin/LSST, Euclid and CSST).


\section*{Acknowledgements}
RL acknowledge the science research grants from the China Manned Space Project (No CMS-CSST-2021-B01,CMS-CSST-2021-A01). NRN acknowledge financial support from the “One hundred top talent program of Sun Yat-sen University” grant N. 71000-18841229.

\bibliographystyle{mnras}
\bibliography{myrefs} 

\begin{appendix}
\section{prior distribution of simulations}
\label{sec:appendix}
In Sect. \ref{sec:training_data}, we have discussed the impact of the distribution of parameters of the mock galaxies used as training and testing sets on the performance of the GaLNets applied to real data. In particular, we have used 
parameter distributions similar to the ones observed in the galaxy sample we expected to analyze, 
e.g. normal distribution for $\log R_{\rm eff}$, and $F$ distribution for $n$. Generally speaking, these prior distributions are unknown,
thus we need to test the impact of inaccurate choices on the final predictions from GaLNets. 
However, in doing this, we should be cautious to adopt unrealistic parameter distributions that might sample a volume in the parameter space that is inaccessible to real galaxies. This can have, as a minimal effect, a degradation of the overall training process because of the loss of useful information in realistic parameter combinations. We stress here, though, that this is the first time such experiments are made to evaluate the response of the Deep Learning in galaxy light profiles using all parameters simultaneously. Hence, we can use this test as a pathfinder for future set-up, although the overall discussion of the prior distribution is beyond the purpose of this paper, but will be tackled in future analyses.

\begin{figure}[htbp]
\centerline{\includegraphics[width=18cm]{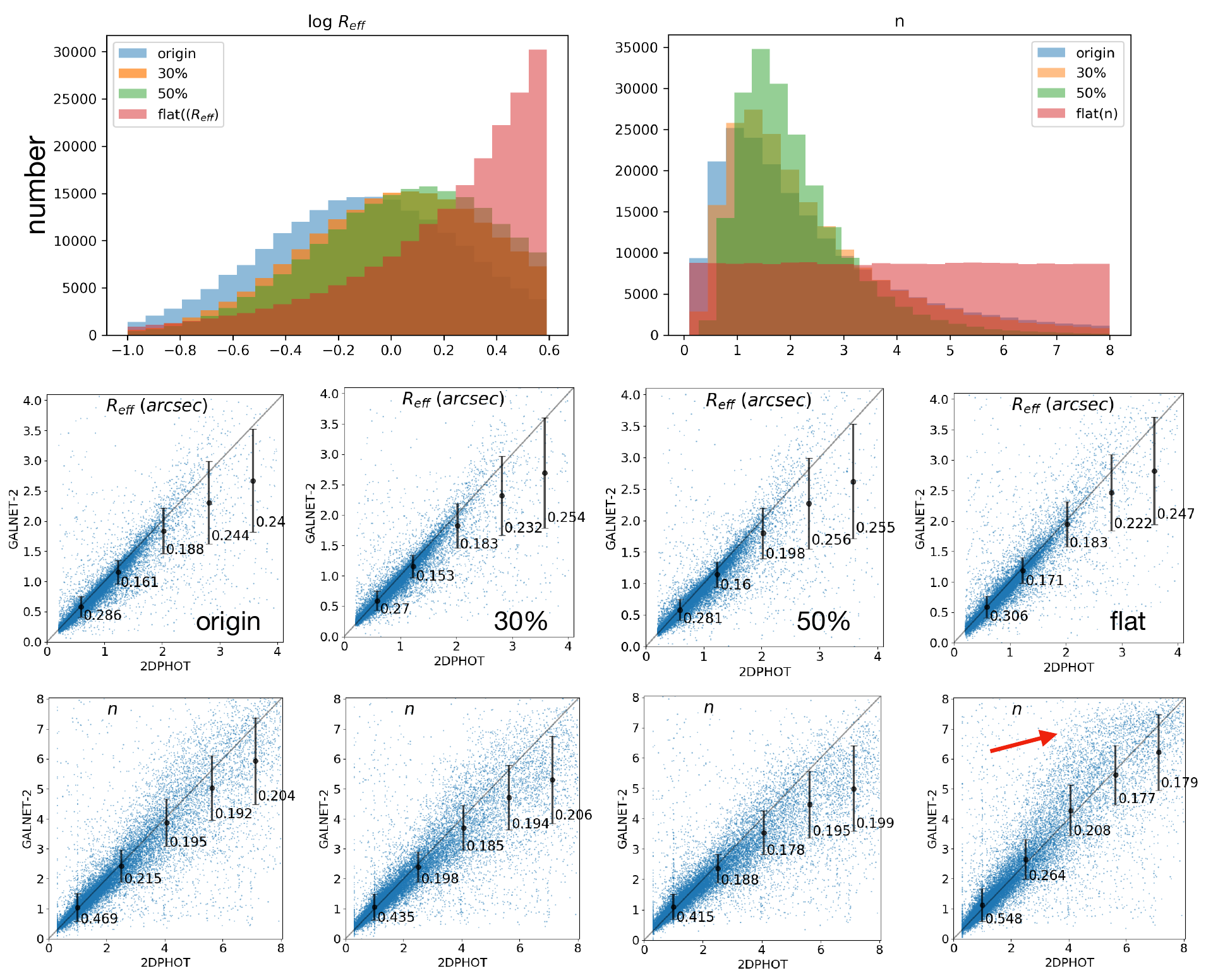}}
\caption{ \rui{Impact of the prior distributions on GaLNet-2 predictions. The two panels in the first row shows the distributions used in simulating the training data. Blue is the original one used in the main paper, orange and green are the ones with 30\% and 50\% peak moving toward to right, red is the one with the flat distribution. The second and third rows show the results obtained with real distribution (origin), 30\% and 50\% peak offset with respect to the original one, and a flat distribution, respectively. The absolute errors bars are also plotted for comparison purpose. The red arrow shows a peculiar feature in the $n$-index predictions (see discussion in the text).}}
\label{fig:prior}
\vspace{0.5cm}
\end{figure}

For this first check, besides the specific choice of the adopted distributions, we also consider how a simple variation of the properties of these distributions can impact the final results. In particular, we want to quantify this effect by testing the GaLNets trained 
on a sample of simulated galaxies for which we have perturbed the prior distributions of the main parameters determining their light profiles, namely the $R_{\rm eff}$ and the $n$. This test is meant to reproduce the systematics one might expect if the CNNs are trained over a simulated galaxy sample 
that deviates from real systems.
In particular, for the training data we adopt the following modifications:
\begin{enumerate}
\item[$\bullet$] We have moved the peak of the $\log R_{\rm eff}$ and the $n$ distributions by 30\% and 50\% (see Fig. \ref{fig:prior}).
\item[$\bullet$] We have changed the distributions of $R_{eff}$ and $n$ to flat.
\end{enumerate}
These tests allow us to check the GaLNets' response over a wide variety of reality mismatches, from extreme (flat) to reasonably moderate (30\%) ones, and estimate their impact on the final parameter predictions. 
The results are shown in Fig. \ref{fig:prior} for GaLNet-2. Here, we see the predictions from GaLNet-2 (y-axis) trained on perturbed sample vs. the fitting results obtained by 2DPHOT (x-axis).
We stress that the training samples have the two parameters changed at the same time, so this is the cumulative effect of a prior mismatch for both of the parameters with the same amount of deviation from the true distributions (the perturbed ones). In this respect, these all represent some worse-case scenarios of poor knowledge of the intrinsic galaxy parameters, while in reality the effective radius is known more robustly than the $n-$index.

In Fig. \ref{fig:prior}, we can see that the impact on $R_{\rm eff}$ and $n$ is different. For the former, moving the peak or changing the distribution to flat almost has no obvious impact on the overall performance.
For $n-$index, there are small systematics for small deviations (30\%), at larger values ($n>4$). However, compared with the original one, the scatters do not increase. 
For larger mismatches of the priors (50\%), the systematics of the $n-$index become rather severe. 
Going to the flat distribution, surprisingly, the scatters at larger value ($n>5$) become smaller than that in the original figure and also the mean values seem more aligned to the one-to-one line. However, a peculiar overdensity of estimates systematically above the one-to-one line is evident at $4\lesssim n\lesssim6$, (see the red arrow in Fig. \ref{fig:prior}), which is compensated by a long tail in the bottom of the same line, leaving the mean value in those bins well aligned with the one-to-one line. This feature and the fact that the accuracy and scatter of the estimates in the low-$n$ ($<4$) bins are worse than the original distribution on the first panel on the left, make the flat distribution a reasonable, but not compelling option.  
Finally, we remark that, apart from the $R_{\rm eff}$ and $n$, we also tested the effects of the distribution of other parameters and no obvious changes were found.

To conclude, we believe that a deeper investigation of the prior mismatch is eventually needed, although the first test presented here does not seem to show that there is much room for improvement. Possibly it is worth looking more into the impact of the flat distributions on both the $R_{\rm eff}$ and $n$, possibly masking obvious area of the parameter space that are away from the typical scaling relations of galaxies. Overall, we need to check more into the reasons of the (still) unsatisfactory amount of outliers, that is possibly driving part of the statistical difference between the predictions from different methods. As discussed in \S\ref{sec:future}, this might be due more to the deblending issue than the poor performance of the GaLNets from biased training sets. 

\end{appendix}

\end{document}